\documentclass{sig-alternate-05-2015}
\usepackage{cite,amsmath,amssymb,amsfonts,dsfont,multirow,multicol,epsfig,url,array,makecell,balance,color}
\usepackage[ruled, vlined, linesnumbered]{algorithm2e}
\usepackage{hhline}
\usepackage{array}
\usepackage{enumerate}
\usepackage{enumitem}
\usepackage{times}
\usepackage{booktabs}
\newcommand{\bigcell}[2]{\begin{tabular}{@{}#1@{}}#2\end{tabular}}

\newtheorem{theorem}{Theorem}

\newtheorem{lemma}{Lemma}

\newtheorem{definition}{Definition}

\newtheorem{observation}{Observation}

\def\header{\vspace{2mm} \noindent}
\def\pheader{\vspace{2.5mm} \noindent}

\def\tblcapup{\vspace{0mm}}
\def\tblcapdown{\vspace{2mm}}
\def\tbldown{\vspace{-0mm}}
\def\figcapup{\vspace{-2mm}}
\def\figcapdown{\vspace{-2mm}}

\newcommand{\pushright}[1]{\ifmeasuring@#1\else\omit\hfill$\displaystyle#1$\fi\ignorespaces}
\newcommand{\pushleft}[1]{\ifmeasuring@#1\else\omit$\displaystyle#1$\hfill\fi\ignorespaces}

\def\done{\hspace*{\fill} {$\square$}}

\def\inV{I}

\def\e{\varepsilon}
\def\O{O}
\def\l{\ell}
\def\E{\mathbb{E}}
\def\scw{\sqrt{c}}

\begin{document}
\begin{sloppy}

\title{SLING: A Near-Optimal Index Structure for SimRank}

\numberofauthors{2}
\author{
\alignauthor
Boyu Tian\\
       \affaddr{Shanghai Jiao Tong University}\\
       \affaddr{China}\\
       \email{bytian@umich.edu}
\alignauthor
Xiaokui Xiao\\
       \affaddr{Nanyang Technological University}\\
       \affaddr{Singapore}\\
       \email{xkxiao@ntu.edu.sg}
}

\CopyrightYear{2016}
\setcopyright{acmlicensed}
\conferenceinfo{SIGMOD'16,}{June 26-July 01, 2016, San Francisco, CA,
USA}
\isbn{978-1-4503-3531-7/16/06}\acmPrice{\$15.00}
\doi{http://dx.doi.org/10.1145/2882903.2915243}

\clubpenalty=10000
\widowpenalty = 10000


\maketitle

\begin{abstract}
SimRank is a similarity measure for graph nodes that has numerous applications in practice. Scalable SimRank computation has been the subject of extensive research for more than a decade, and yet, none of the existing solutions can efficiently derive SimRank scores on large graphs with provable accuracy guarantees. In particular, the state-of-the-art solution requires up to a few seconds to compute a SimRank score in million-node graphs, and does not offer any worst-case assurance in terms of the query error.

This paper presents {\em SLING}, an efficient index structure for SimRank computation. {\em SLING} guarantees that each SimRank score returned has at most $\e$ additive error, and it answers any single-pair and single-source SimRank queries in $O(1/\e)$ and $O(n/\e)$ time, respectively. These time complexities are {\em near-optimal}, and are significantly better than the asymptotic bounds of the most recent approach. Furthermore, {\em SLING} requires only $O(n/\e)$ space (which is also near-optimal in an asymptotic sense) and $O(m/\e + n\log \frac{n}{\delta}/\e^2)$ pre-computation time, where $\delta$ is the failure probability of the preprocessing algorithm. We experimentally evaluate {\em SLING} with a variety of real-world graphs with up to several millions of nodes. Our results demonstrate that {\em SLING} is up to $10000$ times (resp.\ $110$ times) faster than competing methods for single-pair (resp.\ single-source) SimRank queries, at the cost of higher space overheads.
\end{abstract}

\vspace{-2mm}
\section{Introduction} \label{sec:intro}

Assessing the similarity of nodes based on graph topology is an important problem with numerous applications, including social network analysis \cite{NK07}, web mining \cite{Jin11}, collaborative filtering \cite{AMC08}, natural language processing \cite{RotheS14}, and spam detection \cite{SH11}. A number of similarity measures have been proposed, among which {\em SimRank} \cite{JW02} is one of the most well-adopted. The formulation of SimRank is based on two intuitive arguments:
\begin{itemize}[topsep = 5pt, parsep = 5pt, itemsep = 0pt, leftmargin=18pt]
\item A node should have the maximum similarity to itself;

\item The similarity between two different nodes can be measured by the average similarity between the two nodes' neighbors.
\end{itemize}
Formally, the SimRank score of two nodes $v_i$ and $v_j$ is defined as:
\\

\begin{equation} \label{eqn:def-sim}
s(v_i, v_j) =
\begin{cases}
1, & \text{if $v_i = v_j$}\\[2mm]
{\displaystyle \frac{c}{|\inV(v_i)| \cdot |\inV(v_j)|} \sum_{a \in \inV(v_i), b \in \inV(v_j)} s(a, b)}, & \text{otherwise}
\end{cases}
\end{equation}
where $\inV(v)$ denotes the set of in-neighbors of a node $v$, and $c \in (0, 1)$ is a decay factor typically set to $0.6$ or $0.8$ \cite{JW02,LVGT10}. Previous work \cite{NK07,Jin11,AMC08,RotheS14,SH11,FR05,Lin12,YuM15a,ZhaoHS09} has applied SimRank (and its variants) to various problem domains, and has demonstrated that it often provides high-quality measurements of node similarity.

\subsection{Motivation}

Despite of the effectiveness of SimRank, computing SimRank scores efficiently on large graphs is a challenging task, and has been the subject of extensive research for more than a decade. In particular, Jeh and Widom \cite{JW02} propose the first SimRank algorithm, which returns the SimRank scores of all pairs of nodes in the input graph $G$. The algorithm incurs prohibitive costs: it requires $\O\!\left(n^2\right)$ space and $\O\!\left(m^2 \log\frac{1}{\e}\right)$ time, where $n$ and $m$ denote the numbers of nodes and edges in $G$, respectively, and $\e$ is the maximum additive error allowed in any SimRank score. Subsequently, Lizorkin et al.\ \cite{LVGT10} improve the time complexity of the algorithm to $\O\!\left(\log \frac{1}{\e} \cdot \min\{nm, n^3/\log n\}\right)$, which is further improved to $\O\!\left(\log \frac{1}{\e} \cdot \min\{nm, n^\omega\}\right)$ by Yu et al.\ \cite{YZL12}, where $\omega \approx 2.373$. However, the space complexity of the algorithm remains $\O\!\left(n^2\right)$, as is inherent in any algorithm that computes {\em all-pair} SimRank scores.

Fogaras and R{\'{a}}cz \cite{FR05} present the first study on {\em single-pair} SimRank computation, and propose a Monte-Carlo method that requires $\O\!\left(n \log\frac{1}{\delta} /\e^2\right)$ pre-computation time and space. The method returns the SimRank score of any node pair in $\O\!\left(\log\frac{1}{\delta} /\e^2\right)$ time, where $\delta$ is the failure probability of the Monte-Carlo method. Subsequently, Li et al.\ \cite{LLYHD10} propose a deterministic algorithm for single-pair SimRank queries; it has the same time complexity with Jeh and Widom's solution \cite{JW02}, but provides much better practical efficiency. However, existing work \cite{MKK14} show that neither Li et al.'s \cite{LLYHD10} nor Fogaras and R{\'{a}}cz's solution \cite{FR05} is able to handle million-node graphs in reasonable time and space. There is a line of research \cite{FNSO13,He10,Yu13,Li10,Yu14,YuM15b} that attempts to mitigate this efficiency issue based on an alternative formulation of SimRank, but the formulation is shown to be {\em incorrect} \cite{KMK14}, in that it does not return the same SimRank scores as defined in Equation~\eqref{eqn:def-sim}.

\begin{table*} [!t]
\centering
\renewcommand{\arraystretch}{1.5}
\begin{small}
\tblcapup
\caption{Comparison of SimRank computation methods with at most $\boldsymbol{\e}$ additive error and at least $\boldsymbol{1 - \delta}$ success probability.}\label{tbl:intro-compare}
\tblcapdown
 \begin{tabular} {|c|c|c|c|c|} \hline
   \multirow{2}{*}{{\bf Algorithm}}       &   \multicolumn{2}{c|}{{\bf Query Time}} &  \multirow{2}{*}{\bf Space Overhead} & \multirow{2}{*}{\bf Preprocessing Time}\\ \cline{2-3}
   & {\bf Single Pair} & {\bf Single Source} &   & \\ \hline
   Fogaras and R{\'{a}}cz \cite{FR05} & $\O\!\left(\log\frac{1}{\e}\log\frac{n}{\delta}/\e^2\right)$ & $\O\!\left(n \log\frac{1}{\e} \log\frac{n}{\delta} /\e^2\right)$ & $\O\!\left(n \log\frac{1}{\e}\log\frac{n}{\delta} /\e^2\right)$ & $\O\!\left(n \log\frac{1}{\e}\log\frac{n}{\delta} /\e^2\right)$ \\ \hline
   \bigcell{c}{Maehara et al.\ \cite{MKK14} \vspace{-1.5mm} \\  (under heuristic assumptions)} & $\O\!\left(m \log\frac{1}{\e}\right)$  & $\O\!\left(m \log^2\frac{1}{\e}\right)$ & $\O(n + m)$ & no formal result \\ \hline

    \multirow{2}{*}{this paper} & \multirow{2}{*}{$\O(1/\e)$} & $\O(n/\e)$ (Algorithm~\ref{alg:single-pair}) & \multirow{2}{*}{$\O(n/\e)$} & \multirow{2}{*}{$\O\!\left(m/\e + n \log\frac{n}{\delta} / \e^2\right)$} \\ \cline{3-3}
     & & $\O\!\left(m \log^2\frac{1}{\e}\right)$ (Algorithm~\ref{alg:single-source}) & & \\ \hline
   lower bound & $\Omega(1)$ & $\Omega(n)$ & $\Omega(n)$ & - \\ \hline
 \end{tabular}
\end{small}
\tbldown
\vspace{-3mm}
\end{table*}

The most recent approach to SimRank computation is the {\em linearization} technique \cite{MKK14} by Maehara et al., which is shown to considerably outperform existing solutions in terms of efficiency and scalability. Nevertheless, it still requires up to a few seconds to answer a single-pair SimRank query on sizable graphs, which is inadequate for large-scale applications. More importantly, the technique is unable to provide any worst-case guarantee in terms of query accuracy. In particular, the technique has a preprocessing step that requires solving a system $L$ of linear equations; assuming that the solution to $L$ is {\em exact}, Maehara et al.\ \cite{MKK14} show that the technique can ensure $\e$ worst-case query error, and can answer any single-pair and single-source SimRank queries in $\O\!\left(m \log \frac{1}{\e}\right)$ and $\O\!\left(m \log^2 \frac{1}{\e}\right)$ time, respectively. (A single-source SimRank query from a node $v_i$ asks for the SimRank score between $v_i$ and every other node.) Unfortunately, as we discuss in Section~\ref{sec:exist-linear}, the linearization technique cannot precisely solve $L$, nor can it offer non-trivial guarantees in terms of the query errors incurred by the imprecision of $L$'s solution. Consequently, the technique in \cite{MKK14} only provides heuristic solutions to SimRank computation. In summary, after more than tens years of research on SimRank, there is still no solution for efficient SimRank computation on large graphs with provable accuracy guarantees.

\subsection{Contributions and Organization}

This paper presents {\em SLING ({\underline{S}}imRank via {\underline{L}}ocal Updates and Sampl{\underline{ing}})}, an efficient index structure for SimRank computation. {\em SLING} guarantees that each SimRank score returned has at most $\e$ additive error, and answers any single-pair and single-source SimRank queries in $O(1/\e)$ and $O(n/\e)$ time, respectively. These time complexities are {\em near-optimal}, since any SimRank method requires $\Omega(1)$ (resp.\ $\Omega(n)$) time to output the result of any single-pair (resp.\ single-source) query. In addition, they are significantly better than the asymptotic bounds of the 
states of the art (including Maehara et al.'s technique \cite{MKK14} under their heuristic assumptions), as we show in Table~\ref{tbl:intro-compare}. Furthermore, {\em SLING} requires only $O(n/\e)$ space (which is also near-optimal in an asymptotic sense) and $O(m/\e + n\log \frac{n}{\delta})$ pre-computation time, where $\delta$ is the failure probability of the preprocessing algorithm.

Apart from its superior asymptotic bounds, {\em SLING} also incorporates several optimization techniques to enhance its practical performance. In particular, we show that its preprocessing algorithm can be improved with a technique that estimates the expectation of a Bernoulli variable using an {\em asymptotically optimal} number of samples. Additionally, its space consumption can be heuristically reduced without affecting its theoretical guarantees, while its empirical efficiency for single-source SimRank queries can be considerably improved, at the cost of a slight increase in its query time complexity. Last but not least, its construction algorithms can be easily parallelized, and it can efficiently process queries even when its index structure does not fit in the main memory.

We experimentally evaluate {\em SLING} with a variety of real-world graphs with up to several millions of nodes, and show that it significantly outperforms the 
the states of the art in terms of query efficiency. Specifically, {\em SLING} requires at most $2.3$ milliseconds to process a single-pair SimRank query on our datasets, and is up to $10000$ times faster than the linearization method \cite{MKK14}. To our knowledge, this is the first result in the literature that demonstrates millisecond-scale query time for single-pair SimRank computation on million-node graphs. For single-source SimRank queries, {\em SLING} is up to $110$ times more efficient than the linearization method. As a tradeoff, {\em SLING} incurs larger space overheads than the linearization method, but it is a still much more favorable choice in the common scenario where query time and accuracy (instead of space consumption) are the main concern.

The remainder of the paper is organized as follows. Section~\ref{sec:def} defines the problem that we study. Section~\ref{sec:exist} discusses the major existing methods for SimRank computation. Section~\ref{sec:our} presents the {\em SLING} index, with a focus on single-pair queries. Section~\ref{sec:opt} proposes techniques to optimize the practical performance of {\em SLING}. Section~\ref{sec:ext} details how {\em SLING} supports single-source queries. Section~\ref{sec:exp} experimentally evaluates {\em SLING} against the stats of the art 

\section{Preliminaries} \label{sec:def}

Let $G$ be a directed and unweighted graph with $n$ nodes and $m$ edges. We aim to construct an index structure on $G$ to support {\em single-pair} and {\em single-source} SimRank queries, which are defined as follows:
\begin{itemize}[topsep = 6pt, parsep = 6pt, itemsep = 0pt, leftmargin=18pt]
\item A single-pair SimRank query takes as input two nodes $u$ and $v$ in $G$, and returns their SimRank score $s(u, v)$ (see Equation~\ref{eqn:def-sim}).

\item A single-source SimRank query takes as input a node $u$, and returns $s(u, v)$ for each node $v$ in $G$.

\end{itemize}
Following previous work \cite{LVGT10,YuM15a,MKK14,FR05}, we allow an additive error of at most $\e \in (0, 1)$ in each SimRank score returned for any SimRank query.

For ease of exposition, we focus on single-pair SimRank queries in Sections \ref{sec:exist}-\ref{sec:opt}, and then discuss single-source queries in Section~\ref{sec:ext}. Table~\ref{tbl:def-notation} shows the notations frequently used in the paper. Unless otherwise specified, all logarithms in this paper are to base $e$.

\begin{table} [t]
\centering
\renewcommand{\arraystretch}{1.3}
\begin{small}
\tblcapup
\caption{Table of notations.}\label{tbl:def-notation}
\tblcapdown
 \begin{tabular} {|l|p{2.5in}|} \hline
   {\bf Notation}       &   {\bf Description}                                       \\ \hline
   $G$         &   the input graph                                                  \\ \hline
   $n, m$      &   the numbers of nodes and edges in $G$                            \\ \hline
   $v_i$        &   the $i$-th node in $G$                                          \\  \hline
   $I(v)$       &   the set of in-neighbors of a node $v$ in $G$                           \\ \hline
   $s(v_i, v_j)$      &   the SimRank score of two nodes $v_i$ and $v_j$ in $G$     \\ \hline
   $c$          &   the decay factor in the definition of SimRank                   \\ \hline
   $\e$         &   the maximum additive error allowed in a SimRank score           \\ \hline
   $\delta$     &   the failure probability of a Monte-Carlo algorithm              \\ \hline
    $M(i, j)$   & the entry on the $i$-th row and $j$-th column of a matrix $M$   \\ \hline
   $d_k$        &   the correction factor for node $v_k$                            \\ \hline
   $h^{\l}(v_i, v_j)$ & the hitting probability (HP) from node $v_i$ to node $v_j$ at step $\l$ (see Section~\ref{sec:our-overview}) \\ \hline
 \end{tabular}
\end{small}
\end{table}

\section{Analysis of Existing Methods} \label{sec:exist}

This section revisits the three major approaches to SimRank computation: the {\em power method} \cite{JW02}, the {\em Monte Carlo method} \cite{FR05}, and the {\em linearization method} \cite{MKK14,MKK15,KMK14,YuM15a}. The asymptotic performance of the Monte Carlo method and the linearization method has been studied in literature, but to our knowledge, there is no formal analysis regarding their space and time complexities when ensuring $\e$ worst-case errors. We remedy this issue with detailed discussions on each method's asymptotic bounds and limitations.

\subsection{The Power Method} \label{sec:exist-power}

The power method \cite{JW02} is an iterative method for computing the SimRank scores of all pairs of nodes in an input graph. The method uses a $n \times n$ matrix $S$, where the element $S(i, j)$ on the $i$-th row and $j$-th column ($i, j \in [1, n]$) denotes the SimRank score of the $i$-th node $v_i$ and $j$-th node $v_j$. Initially, the method sets
$$S(i, j) =
\begin{cases}
1, & \textrm{if $i = j$} \\
0, & \textrm{otherwise}
\end{cases}$$
After that, in the $t$-th ($t \ge 1$) iteration, the method updates $S$ based on the following equation:
$$S(i,j) =
\begin{cases}
1, & \text{if $i = j$}\\[2mm]
{\displaystyle \frac{c}{|\inV(v_i)|  |\inV(v_j)|} \sum_{v_k \in \inV(v_i), v_\l \in \inV(v_j)} S(k, \l)}, & \text{otherwise}
\end{cases}
$$
Let $S^{(t)}$ denote the version of $S$ right after the $t$-th iteration. Lizorkin et al.\ \cite{LVGT10} establish the following connection between $t$ and the errors in the SimRank scores in $S^{(t)}$:
\begin{lemma}[\cite{LVGT10}]\label{lmm:exist-power}
If ${t \ge \log_c (\e \cdot (1-c)) - 1}$, then for any $i, j \in [1, n]$, we have ${\left|S^{(t)}(i, j) - s(v_i, v_j)\right| \le \e}$ . \done
\end{lemma}

Based on Lemma~\ref{lmm:exist-power} and the fact that each iteration of the power method takes $O\!\left(m^2\right)$ time, we conclude that the power method runs in $\O\!\left(m^2 \log \frac{1}{\e}\right)$ time when ensuring $\e$ worst-case error. In addition, it requires $\O\!\left(n^2\right)$ space (caused by $S$). These large complexities in time and space make the power method only applicable on small graphs.

\subsection{The Monte Carlo Method} \label{sec:exist-MC}

The Monte Carlo method \cite{FR05} is motivated by an alternative definition of SimRank scores \cite{JW02} that utilizes the concept of {\em reverse random walks}. Given a node $w_0$ in $G$, a reverse random walk from $w_0$ is a sequence of nodes $W = \langle w_0, w_1, w_2, \ldots \rangle$, such that $w_{i+1}$ ($i \ge 0$) is selected uniformly at random from the in-neighbors of $w_i$. We refer to $w_i$ as the $i$-th {\em step} of $W$.

Suppose that we have two reverse random walks $W_i$ and $W_j$ that start from two nodes $v_i$ and $v_j$, respectively, and they {\em first meet} at the $\tau$-th step. That is, the $\tau$-th steps of $W_i$ and $W_j$ are identical, but for any $\l \in [0, \tau)$, the $\l$-th step of $W_i$ differs from the $\l$-th step of $W_j$. Jeh and Widom \cite{JW02} establishes the following connection between $\tau$ and the SimRank score of $v_i$ and $v_j$:
\begin{equation} \label{eqn:exist-MC-walk}
s(v_i, v_j) = \E[c^\tau],
\end{equation}
where $\E[\cdot]$ denotes the expectation of a random variable.

Based on Equation~\eqref{eqn:exist-MC-walk}, the Monte Carlo method \cite{FR05} pre-computes a set $\mathcal{W}_i$ of reverse random walks from each node $v_i$ in $G$, such that (i) each set $\mathcal{W}_i$ has the same number $n_w$ of walks, and (ii) each walk in $\mathcal{W}_i$ is truncated at step $t$, i.e., the nodes after the $t$-th step are omitted. (This truncation is necessary to ensure that the walk is computed efficiently.) Then, given two nodes $v_i$ and $v_j$, the method estimates their SimRank score as
\begin{equation*}
\hat{s}(v_i, v_j) = \frac{1}{n_w} \sum_{\l = 0}^{n_w} c^{\tau_\l},
\end{equation*}
where $\tau_\l$ denotes the step at which the $\l$-th walk in $\mathcal{W}_i$ first meets with the $\l$-th walk in $\mathcal{W}_j$. Fogaras and R{\'{a}}cz \cite{FR05} show that, with at least $1 - 2\exp(-\frac{6}{7} n_w \e^2)$ probability,
\begin{equation} \label{eqn:exist-MC-prob}
\big|\hat{s}(v_i, v_j) - \E\left[\hat{s}(v_i, v_j)\right]\big| \le \e.
\end{equation}
However, we note that $\E\left[\hat{s}(v_i, v_j)\right] \ne s(v_i, v_j)$, due to the truncation imposed on the reverse random walks in $W_i$ and $W_j$. To address this issue, we present the following inequality:
\begin{align} \label{eqn:exist-MC-error}
 \Big|\E\left[s(v_i, v_j)\right] - \hat{s}(v_i, v_j)\Big| & = \Big| \E\big[c^\tau\big] - \Pr[\tau \le t] \cdot \E\big[c^\tau \mid \tau \le t\big]  \Big| \nonumber \\
& {} = \big| Pr[\tau > t \big] \cdot \E[c^\tau \mid \tau > t]| \nonumber \\
& {} \le c^{t+1}
\end{align}

By Equations \eqref{eqn:exist-MC-prob} and \eqref{eqn:exist-MC-error} and the union bound, it can be verified that when $t > \log_c \frac{\e}{2}$ and $n_w \ge \frac{14}{3 \e^2} \left(\log \frac{2}{\delta} + 2 \log n\right)$,
$$\big| \hat{s}(v_i, v_j) - s(v_i, v_j) \big| \le \e$$
holds for all pairs of $v_i$ and $v_j$ with at least $1-\delta$ probability. In that case, the space and preprocessing time complexities of the Monte Carlo method are both $O(n_w \cdot t) = O\!\left(\frac{n}{\e^2} \log\frac{1}{\e} \log\frac{n}{\delta}\right)$. In addition, the method takes $O\!\left(\frac{1}{\e^2} \log\frac{1}{\e} \log\frac{n}{\delta}\right)$ time to answer a single-pair SimRank query, and $O\!\left(\frac{n}{\e^2} \log\frac{1}{\e} \log\frac{n}{\delta}\right)$ time to process a single-source SimRank query. These space and time complexities are rather unfavorable under typical settings of $\e$ in practice (e.g., $\e = 0.01$). Fogaras and R{\'{a}}cz \cite{FR05} alleviate this issue with a {\em coupling technique}, which improves the practical performance of the Monte Carlo method in terms of pre-computation time and space consumption. Nevertheless, the method still incurs significant overheads, due to which it is unable to handle graphs with over one million nodes, as we show in Section~\ref{sec:exp}.

\subsection{The Linearization Method} \label{sec:exist-linear}

Let $S$ and $P$ be two $n\times n$ matrices, with $S(i, j) = s(v_i, v_j)$ and
\begin{equation} \label{eqn:exist-linear-transit}
P(i, j) =
\begin{cases}
1/|I(v_j)|, & \textrm{if $v_i \in  I(v_j)$} \\
0, & \textrm{otherwise}
\end{cases}
\end{equation}
Yu et al.\ \cite{YZL12} show that Equation~\eqref{eqn:def-sim} (i.e., the definition of SimRank) can be rewritten as
\begin{equation} \label{eqn:def-sim-mat}
S = (cP^{\top}SP) \lor I,
\end{equation}
where $I$ is an $n \times n$ identity matrix, $P^\top$ is the transpose of $P$, and $\lor$ is the {\em element-wise maximum} operator, i.e., ${(A \lor B)(i,j) = \max\{A(i,j), B(i,j)\}}$ for any two matrices $A$ and $B$ and any $i, j$. 

Maehara et al.\ \cite{MKK14} point out that solving Equation~\eqref{eqn:def-sim-mat} is difficult since it is a {\em non-linear} problem due to the $\lor$ operator. To circumvent this difficulty, they prove that there exists a $n \times n$ diagonal matrix $D$ (referred to as the {\em diagonal correction matrix}), such that
\begin{equation} \label{eqn:def-lsim}
S = cP^{\top}SP + D.
\end{equation}
Furthermore, once $D$ is given, one can uniquely derive $S$ based on the following lemma by Maehara et al.\ \cite{MKK14}:
\begin{lemma}[\cite{MKK14}] \label{lmm:exist-lsim}
Given the diagonal correction matrix $D$,
\begin{equation} \label{eqn:def-lsim2}
S = \sum_{\l = 0}^{+\infty} c^\l \left(P^\l\right)^\top D P^\l,
\end{equation}
where $P^\l$ denotes the $\l$-th power of $P$. \done
\end{lemma}

Given Lemma~\ref{lmm:exist-lsim}, Maehara et al.\ \cite{MKK14} propose the linearization method, which pre-computes $D$ and then uses it to answer SimRank queries based on Equation~\eqref{eqn:def-lsim2}. In particular, for any two nodes $v_i$ and $v_j$, Equation~\eqref{eqn:def-lsim2} leads to
\begin{equation} \label{eqn:def-lsim-single-pair}
s(v_i, v_j) = \sum_{\l = 0}^{+\infty} c^\l \left(P^\l \cdot \vec{e_i}\right)^\top D \left(P^\l \cdot \vec{e_j}\right),
\end{equation}
where $\vec{e_k}$ denotes a $n$-element column vector where the $k$-th element equals $1$ and all other elements equal $0$. To avoid the infinite series in Equation~\eqref{eqn:def-lsim-single-pair}, the linearization method approximates $s(v_i, v_j)$ with
\begin{equation} \label{eqn:def-lsim-single-pair2}
\tilde{s}(v_i, v_j) = \sum_{\l = 0}^{t} c^\l \left(P^\l \cdot \vec{e_i}\right)^\top D \left(P^\l \cdot \vec{e_j}\right),
\end{equation}
which can be computed in $O(m \cdot t)$ time. It can be shown that if $D$ is precise and ${t \ge \log_c (\e \cdot (1-c)) - 1}$, then
\begin{equation} \label{eqn:exist-lsim-error}
\big| \tilde{s}(v_i, v_j) - s(v_i, v_j) \big| \le \e.
\end{equation}
Therefore, given an exact $D$, the linearization method answers any single-pair SimRank query in $O(m \log\frac{1}{\e})$ time. With a slight modification of Equation~\ref{eqn:def-lsim-single-pair2}, the method can also process any single-source SimRank query in $O(m \log^2\frac{1}{\e})$ time.

Unfortunately, the linearization method do not precisely derive $D$, due to which the above time complexities does not hold in general. Specifically, Maehara et al.\ \cite{MKK14} formulate $D$ as the solution to a linear system, and propose to solve an {\em approximate} version of the system to derive an estimation $\widetilde{D}$ of $D$. However, there is no formal analysis on the errors in $\widetilde{D}$ and their effects on the accuracy of SimRank computation. In addition, the technique used to solve the approximate linear system does not guarantee to {\em converge}, i.e., it may not return $\widetilde{D}$ in bounded time. Furthermore, even if the technique does converge, its time complexity relies on a parameter that is unknown in advance, and may even dominate $n$, $m$, and $1/\e$. This makes it rather difficult to analyze the pre-computation time of the linearization method. We refer interested readers to Appendix~\ref{apnx:linear} for detailed discussions on these issues.

In summary, the linearization method by Maehara et al.\ \cite{MKK14} does not guarantee $\e$ worst-case error in each SimRank score returned, and there is no non-trivial bound on its preprocessing time. This problem is partially addressed in recent work \cite{YuM15a} by Yu and McCann, who propose a variant of the linearization method that does not pre-compute the diagonal correction matrix $D$, but implicitly derives $D$ during query processing. Yu and McCann's technique is able to ensure $\e$ worst-case error in SimRank computation, but as a trade-off, it requires $O\!\left(mn\log\frac{1}{\e}\right)$ time to answer a single-pair SimRank query, which renders it inapplicable on any sizable graph.

\section{Our Solution} \label{sec:our}

This section presents our {\em SLING} index for SimRank queries. {\em SLING} is based on a new interpretation of SimRank scores, which we clarify in Section~\ref{sec:our-interpret}. After that, Sections \ref{sec:our-D}-\ref{sec:our-query} provide details of {\em SLING} and analyze its theoretical guarantees.

\subsection{New Interpretation of SimRank} \label{sec:our-interpret}

Let $c$ be the decay factor in the definition of SimRank (see Equation~\eqref{eqn:def-sim}). Suppose that we perform a reverse random walk from any node $u$ in $G$, such that
\begin{itemize}[topsep = 6pt, parsep = 6pt, itemsep = 0pt, leftmargin=18pt]
\item At each step of the walk, we stop with $1 - \sqrt{c}$ probability;

\item With the other $\sqrt{c}$ probability, we inspect the in-neighbors of the node at the current step, and select one of them uniformly at random as the next step.
\end{itemize}
We refer to such a reverse random walk as a {\em $\scw$-walk} from $u$. In addition, we say that two $\scw$-walks {\em meet}, if for a certain $\l \ge 0$, the $\l$-th steps of the two walks are identical. (Note the $0$-th step of a $\scw$-walk is its starting node.) The following lemma shows an interesting connection between $\scw$-walks and SimRank.

\begin{lemma} \label{lmm:our-scwalk}
Let $W_i$ and $W_j$ be two $\scw$-walks from two nodes $v_i$ and $v_j$, respectively. Then, $s(v_i, v_j)$ equals the probability that $W_i$ and $W_j$ meet. \done
\end{lemma}

The above formulation of SimRank is similar in spirit to the one used in the Monte Carlo method \cite{FR05} (see Section~\ref{sec:exist-MC}), but differs in one crucial aspect: each $\scw$-walk in our formulation has an expected length of $\frac{1}{1-\scw}$, whereas each reverse random walk in the previous formulation is infinite. As a consequence, if we are to estimate $s(v_i, v_j)$ using a sample set of $\scw$-walks from $v_i$ and $v_j$, we do not need to truncate any $\scw$-walk for efficiency; in contrast, the Monte Carlo method \cite{FR05} must trim each reverse random walk to trade estimation accuracy for bounded computation time. In fact, if we incorporate $\scw$-walks into the Monte Carlo method, then its query time complexities are immediately improved by a factor of $\log \frac{1}{\e}$. Nonetheless, the space and time overheads of this revised method still leave much room for improvement, since it requires $O(\log\frac{n}{\delta}/\e^2)$ $\scw$-walks for each node, where $\delta$ is the upper bound on the method's failure probability. This motivates us to develop the {\em SLING} method for more efficient SimRank computation, which we elaborate in the following sections.

\subsection{Key Idea of {\large \bf \em SLING}} \label{sec:our-overview}

Let $h^{(\l)}(v_a, v_b)$ denote the probability that a $\scw$-walk from $v_a$ arrives at $v_b$ in its $\l$-th step. We refer to $h^{(\l)}(v_a, v_b)$ as the {\em hitting probability ({\bf \em HP})} from $v_a$ to $v_b$ at step $\l$. Observe that, for any two $\scw$-walks $W_i$ and $W_j$ from two nodes $v_i$ and $v_j$, respectively, the probability that they meet at $v_k$ at the $\l$-th step is $$h^{(\l)}(v_i, v_k) \cdot h^{(\l)}(v_j, v_k).$$
Since $s(v_i, v_j)$ equals the probability that $W_i$ and $W_j$ meet, one may attempt to compute $s(v_i, v_j)$ by taking the the probability that $W_i$ and $W_j$ meet over all combinations of meeting nodes and meeting steps, i.e.,
\begin{equation} \label{eqn:our-lsim-wrong}
s^*(v_i, v_j) = \sum_{\l=0}^{+\infty} \sum_{k = 1}^n \left(h^{(\l)}(v_i, v_k) \cdot h^{(\l)}(v_j, v_k)\right).
\end{equation}
However, this formulation is incorrect, because the events that ``$W_i$ and $W_j$ meet at node $v_x$ at step $\l$'' and ``$W_i$ and $W_j$ meet at node $v_y$ at step $\l' > \l$'' are {\em not} mutually exclusive. For example, assume that $v_i = v_j$, and $v_i$ has only in-neighbor $v_k$. In that case, $W_i$ and $W_j$ have $100\%$ probability to meet at $v_i$ at the $0$-th step, and a non-zero probability to meet at $v_k$ at the first step. This leads to $s^*(v_i, v_j) > 1$, whereas $s(v_i, v_j) = 1$ by definition.

Interestingly, Equation~\eqref{eqn:our-lsim-wrong} can be fixed if we substitute ${h^{(\l)}(v_i, v_k) \cdot h^{(\l)}(v_j, v_k)}$ with the probability of the event that ``$W_i$ and $W_j$ meet at $v_k$ at step $\l$, {\em but never meet again afterwards}''. To explain this, observe that the above event indicates that $W_i$ and $W_j$ {\em last meet} at $v_k$ at step $\l$. If we change $v_k$ (resp.\ $\l$) in the event, then $W_i$ and $W_j$ should last meet at a different node (resp.\ step), in which case the changed event and the original one are mutually exclusive. Based on this observation, the following lemma presents a remedy to Equaiton~\eqref{eqn:our-lsim-wrong}.
\begin{lemma} \label{lmm:our-lsim}
Let $d_k$ be the probability that two $\scw$-walks from node $v_k$ do not meet each other after the $0$-th step. Then, for any two nodes $v_i$ and $v_j$,
\begin{equation} \label{eqn:our-lsim}
s(v_i, v_j) = \sum_{\l = 0}^\infty \sum_{k = 1}^n \left(h^{(\l)}(v_i, v_k) \cdot d_k \cdot h^{(\l)}(v_j, v_k)\right).
\end{equation}
\end{lemma}
In what follows, we refer to $d_k$ as the {\em correction factor} for $v_k$.

Based on Lemma~\ref{lmm:our-lsim}, we propose to pre-compute approximate versions of $d_k$ and HPs $h^{(\l)}(v_i, v_k)$, and then use them to estimate SimRank scores based on Equation~\eqref{lmm:our-lsim}. The immediate problem here is that there exists an infinite number of HPs $h^{(\l)}(v_i, v_k)$ to approximate, since we need to consider all $\l \ge 0$. However, we observe that if we allow an additive error in the approximate values, then most of the HPs can be estimated as zero and be omitted. In particular, we have the following observation:
\begin{observation} \label{obs:our-num-entry}
For any node $v_i$ and $\l \ge 0$, there exist at most $(\scw)^\l / \e_h$ nodes $v_k$ such that $h^{(\l)}(v_j, v_k) \ge \e_h$. \done
\end{observation}

To understand this, recall that each $\scw$-walk has only $(\scw)^\l$ probability to {\em not} stop before the $\l$-th step, i.e.,
$${\sum_{k = 1}^n h^{(\l)}(v_j, v_k) = (\scw)^\l}.$$
Therefore, at most $(\scw)^\l/\e_h$ of the HPs at step $\l$ can be larger than $\e_h$. Even if we take into account all $\l \ge 0$, the total number of HPs above $\e_h$ is only
$$\sum_{\l = 0}^{+\infty} (\scw)^\l/\e_h = O(1/\e_h).$$
In other words, we only need to retain a constant number of HPs for each node, if we permit a constant additive error in each HP.

Based on the above analysis, we propose the {\em SLING} index, which pre-computes an approximate version $\tilde{d_k}$ of each correction factor $d_k$, as well as a constant-size set $H(v_i)$ of approximate HPs for each node $v_i$. To derive the SimRank score of two nodes $v_i$ and $v_j$, {\em SLING} first retrieves $\tilde{d_k}$, $H(v_i)$, and $H(v_j)$, and then estimates $s(v_i, v_j)$ in constant time based on an approximate version of Equation~\eqref{eqn:our-lsim}. The challenge in the design of {\em SLING} is threefold. First, how can we derive an accurate estimation of $\tilde{d_k}$? Second, how can we efficiently construct $H(v_i)$ without iterating over all HPs? Third, how do we ensure that all $\tilde{d_k}$ and $H(v_i)$ can jointly guarantee $\e$ worst-case error in each SimRank score computed? In Sections \ref{sec:our-D}-\ref{sec:our-query}, we elaborate how we address these challenges.

Before we proceed, we note that there is an interesting connection between Lemmas \ref{lmm:exist-lsim} and \ref{lmm:our-lsim}:
\begin{lemma} \label{lmm:our-dk-D}
Let $P$ and $D$ be as in Lemma~\ref{lmm:exist-lsim}, and $d_k$ and $h^{(\l)}(v_i, v_k)$ be as in Lemma~\ref{lmm:our-lsim}. For any $i, k \in [1, n]$, $h^{(\l)}(v_i, v_k) = \left(\sqrt{c}\right)^\l \cdot P(k, i)$, and $d_k$ equals the $k$-th diagonal element in $D$. \done
\end{lemma}
In other words, $h^{(\l)}(v_i, v_k)$ (resp.\ $d_k$) can be regarded as a random-walk-based interpretation of the entries in $P$ (resp.\ diagonal elements in $D$). Therefore, Lemmas \ref{lmm:exist-lsim} and \ref{lmm:our-lsim} are different interpretations of the same result. The main advantage of our new interpretation is that it gives a physical meaning to $d_k$ which, as we show in Section~\ref{sec:our-D}, enables us to devise a simple and rigorous algorithm to estimate $d_k$ to any desired precision. In contrast, the only existing method for approximating $D$ \cite{MKK14} fails to provide any non-trivial guarantees in terms of accuracy and efficiency, as we discuss in Section~\ref{sec:exist-linear}.

\subsection{Estimation of $\large \boldsymbol{d_k}$ } \label{sec:our-D}

Let $W$ and $W'$ be two $\scw$-walks from $v_k$. By definition, ${1 - d_k}$ is the probability that any of the following events occurs:
\begin{enumerate}[topsep = 6pt, parsep = 6pt, itemsep = 0pt, leftmargin=20pt]
\item $W$ and $W'$ meet at the first step.

\item In the first step, $W$ and $W'$ arrive at two different nodes $v_i$ and $v_j$, respectively; but sometime after the first step, $W$ and $W'$ meet.
\end{enumerate}
Note that the above two events are mutually exclusive, and the first event occurs with $\frac{c}{|I(v_k)|}$ probability. For the second event, if we fix a pair of $v_i$ and $v_j$, then the probability that $W$ and $W'$ meet after the first step equals the probability that a $\scw$-walk from $v_i$ meets a $\scw$-walk from $v_j$; by Lemma~\ref{lmm:our-scwalk}, this probability is exactly $s(v_i, v_j)$. Therefore, we have
\begin{align} \label{eqn:our-dk-def}
d_k & = 1 - \frac{c}{|I(v_k)|} - \frac{c}{|I(v_k)|^2} \sum_{\substack{v_i, v_j\in \inV(v_k) \\ v_i \neq v_j}} s(v_i, v_j).
\end{align}
Equation~\eqref{eqn:our-dk-def} indicates that, if we are to estimate $d_k$, it suffices to derive an estimation of
\begin{equation} \label{eqn:our-D-mu}
\mu = \frac{1}{|I(v_k)|^2}\sum_{\substack{v_i, v_j\in \inV(v_k) \land v_i \neq v_j}} s(v_i, v_j)
\end{equation}
by sampling $\scw$-walks from $v_i$ and $v_j$. In particular, as long as $\mu$ is estimated with an error no more than $\e_d/c$, the resulting estimation of $d_k$ would have at most $\e_d$ error. Motivated by this, we propose a sampling method for approximating $d_k$, as shown in Algorithm~\ref{alg:naive-d}.

\begin{algorithm}[t] \label{alg:naive-d}
\begin{small}
\caption{\em A sampling method for estimating $d_k$}
\KwIn {a node $v_k$, an error bound $\e_d$, and a failure probability $\delta_d$}
\KwOut {an estimation version $\tilde{d_k}$ of $d_k$ with at most $\e_d$ error, with at least $1 - \delta_d$ probability}
\BlankLine
Let $n_r = \dfrac{2c^2+c \cdot \e_d}{\varepsilon_d^2} \log{\dfrac{2}{\delta_d}}$\;

Let $cnt = 0$\;

\For {$x = 1, 2, \cdots, n_r$}
{
    Select two nodes $v_i$ and $v_j$ from $I(v_k)$ uniformly at random\;
    \If {$v_i \neq v_j$}
    {
        Generate two $\scw$-walks from $v_i$ and $v_j$, respectively\;
        \If {the two $\scw$-walks meet}
        {
            $cnt = cnt + 1$\;
        }
    }
}
\Return {$\tilde{d_k} = 1 - \dfrac{c}{|\inV(v_i)|} - c \cdot \dfrac{cnt}{n_r}$}\;
\end{small}
\end{algorithm}

In a nutshell, Algorithm~\ref{alg:naive-d} generates $n_r$ pairs of $\scw$-walks, such that each walk starts from a randomly selected node in $I(v_k)$; after that, the algorithm counts the number $cnt$ of pairs that meet at or after the first step; finally, it returns ${\tilde{d_k} = 1 - \frac{c}{|\inV(v_i)|} - c \cdot \frac{cnt}{n_r}}$ as an estimation of $d_k$. By the Chernoff bound (see Appendix~\ref{apnx:concen}) and the properties of $\scw$-walks, we have the following lemma on the theoretical guarantees of Algorithm~\ref{alg:naive-d}.

\begin{lemma} \label{lmm:our-D}
Algorithm~\ref{alg:naive-d} runs in $O\left(\frac{1}{\e_d^2} \log \frac{1}{\delta_d}\right)$ expected time, and returns $\tilde{d_k}$ such that $|\tilde{d_k} - d| \le \e_d$ holds with at least $1 - \delta_d$ probability. \done
\end{lemma}

\begin{algorithm}[t] \label{alg:backward}
\begin{small}
\caption{\em A local update method for constructing $H(v_i)$}
\KwIn{$G$ and a threshold $\theta$}
\KwOut {A set $H(v_i)$ of approximate HPs for each node $v_i$ in $G$}
\BlankLine
Initialize $H(v_i) = \emptyset$ for each node $v_i$\;
\For {each node $v_k$ in $G$}
{
    Initialize a set $R_k = \emptyset$ for storing approximate HPs\;
    Insert $\tilde{h}^{(0)}(v_k, v_k) = 1$ into $R_k$\;
    \For {$\l = 0, 1, 2, \ldots$}
    {
        \For{each $\tilde{h}^{(\l)}(v_x, v_k) \in R_k$}
        {
            \If {$\tilde{h}^{(\l)}(v_x, v_k) \le \theta$}
            {
                remove $\tilde{h}^{(\l)}(v_x, v_k)$ from $R_k$\;
                {\bf continue}\;
            }
            \For{each out-neighbor $v_i$ of $v_x$}
            {
                \If{$\tilde{h}^{(\l)}(v_i, v_k) \notin R_k$}
                {
                    Insert $\tilde{h}^{(\l+1)}(v_i, v_k) = \scw \cdot \frac{\tilde{h}^{(\l)}(v_x, v_k)}{|I(v_i)|}$ into $R_k$\;
                }
                \Else
                {
                    Increase $\tilde{h}^{(\l+1)}(v_i, v_k)$ by $\scw \cdot \frac{\tilde{h}^{(\l)}(v_x, v_k)}{|I(v_i)|}$\;
                }
            }
            \If {$R_k$ does not contain any HP at step $\l+1$}
            {
                {\bf break}\;
            }
        }
    }
    \For {each $\tilde{h}^{(\l)}(v_i, v_k) \in R_k$}
    {
        Insert $\tilde{h}^{(\l)}(v_i, v_k)$ into $H(v_i)$\;
    }
}
\end{small}
\end{algorithm}

\subsection{Construction of $\large \boldsymbol{H(v_i)}$ } \label{sec:our-R}

As mentioned in Section~\ref{sec:our-overview}, we aim to construct a constant-size set $H(v_i)$ for each node $v_i$, such that $H(v_i)$ contains an approximate version $\tilde{h}^{(\l)}(v_i, v_x)$ of each HP ${h}^{(\l)}(v_i, v_x)$ that is sufficiently large. Towards this end, a relatively straightforward solution is to sample a set $\mathcal{W}_i$ of $\scw$-walks from each $v_i$, and then use $\mathcal{W}_i$ to derive approximate HPs. This solution, however, requires $O(1/{\e_h}^2)$ walks in $\mathcal{W}_i$ to ensure that the additive error in each $\tilde{h}^{(\l)}(v_i, v_x)$ is at most $\e_h$, which leads to considerable computation costs when $\e_h$ is small.

Instead of sampling $\scw$-walks, we devise a deterministic method for constructing all $H(v_i)$ in $O(m/{\e_h})$ time while allowing at most $\e_h$ additive error in each approximate HP. The key idea of our method is to utilize the following equation on HPs:
\begin{equation} \label{eqn:our-R-hitting}
h^{(\l+1)}(v_i, v_k) = \dfrac{\sqrt{c}}{|\inV(v_i)|} \sum_{v_x \in I(v_i)} h^{(\l)}(v_x, v_k),
\end{equation}
for any $\l \ge 0$. Intuitively, Equation~\eqref{eqn:our-R-hitting} indicates that once we have derived the HPs to $v_k$ at step $\l$, then we can compute the HPs to $v_k$ at step $\l+1$. Based on this intuition, our method generates approximate HPs to $v_k$ by processing the steps $\l$ in ascending order of $\l$. We note that our method is similar in spirit to the {\em local update} algorithm \cite{ACL06,JehW03,FRCS05} for estimating {\em personalized PageRanks} \cite{JehW03}, and we refer interested readers to Appendix~\ref{apnx:pagerank} for a discussion on the connections between our method and those in \cite{ACL06,JehW03,FRCS05}.

Algorithm~\ref{alg:backward} shows the pseudo-code of our method. Given $G$ and a threshold $\theta$, the algorithm first initializes $H(v_i) = \emptyset$ for each node $v_i$ (Line 1). After that, for each node $v_k$, the algorithm performs a graph traversal from $v_k$ to generates approximate HPs from other nodes to $v_k$. Specifically, for each $v_k$, it first initializes a set $R_k = \emptyset$, and then inserts an HP $\tilde{h}^{(0)}(v_k, v_k) = 1$ into $R_k$, which captures the fact that every $\scw$-walk from $v_k$ has $100\%$ probability to hit $v_k$ itself at the $0$-th step (Lines 3-4). Then, the algorithm enters an iterative process, such that the $\l$-th iteration ($\l \ge 0$) processes the HPs to $v_k$ at step $\l$ that have been inserted into $R_k$.

In particular, in the $\l$-the iteration, the algorithm first identifies the approximate HPs $\tilde{h}^{(\l)}(v_x, v_k)$ in $R_k$ that are at step $\l$, and processes each of them in turn (Lines 6-16). If $\tilde{h}^{(\l)}(v_x, v_k) \le \theta$, then it is removed from $R_k$, i.e., the algorithm omits an approximate HP if it is sufficiently small. Meanwhile, if $\tilde{h}^{(\l)}(v_x, v_k) > \theta$, then the algorithm inspects each out-neighbor $v_i$ of $v_x$, and updates the approximate HP from $v_i$ to $v_k$ at step $\l+1$, according to Equation~\eqref{eqn:our-R-hitting}. After all approximate HPs at step $\l$ are processed, the algorithm terminates the iterative process on $\l$. Finally, the algorithm inserts each $\tilde{h}^{(\l)}(v_i, v_k) \in R$ into $H(v_i)$, after which it proceeds to the next node $v_{k+1}$.

The following lemma states the guarantees of Algorithm~\ref{alg:backward}.
\begin{lemma} \label{lmm:our-R}
Algorithm~\ref{alg:backward} runs in $O(m/\theta)$ time, and constructs a set $H(v_i)$ of approximate HPs for each node $v_i$, such that $|H(v_i)| = O(1/\theta)$. In addition, for each $\tilde{h}^{(\l)}(v_i, v_k) \in H(v_i)$, we have
$$0 \ge \tilde{h}^{(\l)}(v_i, v_k) - h^{(\l)}(v_i, v_k) \ge -\frac{1-(\scw)^\l}{1 - \scw} \cdot \theta.$$
\end{lemma}

\subsection{Query Method and Complexity Analysis} \label{sec:our-query}

Given an approximate correction factor $\tilde{d}_k$ and a set $H(v_k)$ of approximate HPs for each node $v_k$, we estimate the SimRank score between any two nodes $v_i$ and $v_j$ according to a revised version of Equation~\eqref{eqn:our-lsim}:
\begin{equation} \label{eqn:our-lsim-apprx}
\tilde{s}(v_i, v_j) = \sum_{\l = 0}^\infty \sum_{k = 1}^n \left(\tilde{h}^{(\l)}(v_i, v_k) \cdot \tilde{d_k} \cdot \tilde{h}^{(\l)}(v_j, v_k)\right).
\end{equation}
Algorithm~\ref{alg:single-pair} shows the details of our query processing method.

\begin{algorithm}[t] \label{alg:single-pair}
\begin{small}
\caption{\em An algorithm for single-pair SimRank queries}
\KwIn {$\tilde{d}_k$, $H(v_k)$, and two nodes $v_i$ and $v_j$}
\KwOut {An approximate SimRank score $\tilde{s}(v_i, v_j)$}
\BlankLine
Let $\tilde{s}(v_i, v_j) = 0$\;
\For {each $\tilde{h}^{(\l)}(v_i, v_k) \in H(v_i)$}
{
    \If{there exists $\tilde{h}^{(\l)}(v_j, v_k) \in H(v_j)$}
    {
        $\tilde{s}(v_i, v_j) = \tilde{s}(v_i, v_j) + \tilde{h}^{(\l)}(v_i, v_k) \cdot \tilde{d_k} \cdot \tilde{h}^{(\l)}(v_j, v_k)$\;
    }
}
\Return $\tilde{s}(v_i, v_j)$\;
\end{small}
\end{algorithm}

To analyze the accuracy guarantee of Algorithm~\ref{alg:single-pair}, we first present a lemma that quantifies the error in $\tilde{s}(v_i, v_j)$ based on the errors in $\tilde{d_k}$ and $H(v_k)$.
\begin{lemma} \label{lmm:our-error-allowed}
Suppose that $\left|\tilde{d_k} - d_k\right| \le \e_d$ for any $k$, and
$$0 \ge \tilde{h}^{(\l)}(v_k, v_x)- h^{(\l)}(v_k, v_x) \ge -\e^{(\l)}_h,$$
for any $k, x, \l$. Then, we have $\left|\tilde{s}(v_i, v_j) - s(v_i, v_j)\right| \le \e$ if
$$\frac{\e_d}{1-c} + 2 \sum_{\l=0}^{+\infty} \left((\scw)^{\l} \cdot \e^{(\l)}_h\right) \le \e.$$  
\end{lemma}

Combining Lemmas \ref{lmm:our-D}, \ref{lmm:our-R}, and \ref{lmm:our-error-allowed}, we have the following theorem.
\begin{theorem} \label{thm:err-bound}
Suppose that we derive each $\tilde{d_k}$ using Algorithm~\ref{alg:naive-d} with input $\e_d$ and $\delta_d$, and we construct each $H(v_k)$ using Algorithm~\ref{alg:backward} with input $\theta$.
If $\delta_d \le \delta/n$ and
$$\frac{\e_d}{1-c} + \frac{2 \scw}{(1-\scw)(1-c)} \theta \leq \e,$$
then Algorithm~\ref{alg:single-pair} incurs an additive error at most $\e$ in each SimRank score returned, with at least $1 - \delta$ probability. \done
\end{theorem}

By Theorem~\ref{thm:err-bound}, we can ensure $\e$ worst-case error in each SimRank score by setting $\e_d = O(\e)$, $\theta = O(\e)$, and $\delta_d = \delta/n$. In that case, our {\em SLING} index requires $O(m/\e + n \log\frac{n}{\delta})$ pre-computation time and $O(n/\e)$ space, and it answers any single-pair SimRank query in $O(1/\e)$ time. The space (resp.\ query time) complexity of {\em SLING} is only $O(1/\e)$ times larger than the optimal value, since any SimRank method (that ensures $\e$ worst-case error) requires $\Omega(n)$ space for storing the information about all nodes, and takes at least $\Omega(1)$ time to output the result of a single-pair SimRank query.

\section{Optimizations} \label{sec:opt}

This section presents optimization techniques to (i) improve the efficiency of estimating each correction factors $d_k$ (Section~\ref{sec:opt-D}), (ii) reduce the space consumption of {\em SLING} (Section~\ref{sec:opt-R}), (iii) enhance the accuracy of {\em SLING} (Section~\ref{sec:opt-accuracy}), and (iv) incorporate parallel and out-of-core computation into {\em SLING}'s index construction algorithm (Section~\ref{sec:opt-parallel}).

\subsection{Improved Estimation of $\large \boldsymbol{d_k}$} \label{sec:opt-D}

As discussed in Section~\ref{sec:our-D}, Algorithm~\ref{alg:naive-d} generates an approximate correction factor $\tilde{d_k}$ in $O\left(\e_d^{-2} \log \delta_d^{-1}\right)$ expected time, where $\e_d$ is the maximum error allowed in $\tilde{d_k}$, and $\delta_d$ is the failure probability. As the algorithm's time complexity is quadratic to $1/\e_d$, it is not particularly efficient when $\e_d$ is small. This relative inefficiency is caused by the fact the algorithm requires $O\left(\e_d^{-2} \log \delta_d^{-1}\right)$ pairs of $\scw$-walks to estimate the value $\mu$ (in Equation~\ref{eqn:our-D-mu}) with $\e_d/c$ worst-case error.

However, we observe that we can often use a much smaller number of $\scw$-walk pairs to derive an estimation of $\mu$ with at most $\e_d/c$ error. Specifically, by the Chernoff bound (see Appendix~\ref{apnx:concen}), we only need $O\left( (\mu + \e_d) \cdot \e_d^{-2} \log \delta_d^{-1} \right)$ pairs of $\scw$-walks to estimate $\mu$. Apparently, this number is much smaller than $O\left(\e_d^{-2} \log \delta_d^{-1}\right)$ when $\mu \ll 1$ (which is often the case in practice). For example, if $\mu \le \e_d$, then the number of $\scw$-walk pairs required is only $O(\e_d^{-1} \log \delta_d^{-1})$. The main issue here is that we do not know $\mu$ in advance. Nevertheless, if we can derive an upper bound of $\mu$, and we use it to decide an appropriate number of $\scw$-walks needed.

\begin{algorithm}[t] \label{alg:sample-small}
\begin{small}
\caption{\em An improved method for estimating $d_k$}
\KwIn {a node $v_k$, an error bound $\e_d$, and a failure probability $\delta_d$}
\KwOut {an estimation version $\tilde{d_k}$ of $d_k$ with at most $\e_d$ error, with at least $1 - \delta_d$ probability}
\BlankLine
    Let $n_r = \dfrac{14 c}{3\e_d}\log{\dfrac{4}{\delta_d}}$\;
    Let $cnt = 0$\;
    \For {$x = 1, 2, \cdots, n_r$}
    {
        Select two nodes $v_i$ and $v_j$ from $I(v_k)$ uniformly at random\;
        \If {$v_i \neq v_j$}
        {
            Generate two $\scw$-walks from $v_i$ and $v_j$, respectively\;
            \If {the two $\scw$-walks meet}
            {
                $cnt = cnt + 1$\;
            }
        }
    }
    Let $\hat{\mu} = cnt/n_r$\;
    \If {$\hat{\mu} \le \e_d$}
    {
        \Return $\tilde{d_k} = 1 - \dfrac{c}{|\inV(v_i)|} - c \cdot \hat{\mu}$\;
    }
    Let $\mu^* = \hat{\mu} + \sqrt{\hat{\mu} \cdot \e_d}$\;
    Let $n_r^* = \dfrac{2c^2\cdot\mu^* + \frac{2}{3}c \cdot \e_d}{\e_d^2}\log{\dfrac{4}{\delta_d}}$\;
    \For {$x = 1, 2, \cdots, n_r^* - n_r$}
    {
        Select two nodes $v_i$ and $v_j$ from $I(v_k)$ uniformly at random\;
        \If {$v_i \neq v_j$}
        {
            Generate two $\scw$-walks from $v_i$ and $v_j$, respectively\;
            \If {the two $\scw$-walks meet}
            {
                $cnt = cnt + 1$\;
            }
        }
    }
    $\tilde{\mu} = cnt / n_r^*$\;
    \Return $\tilde{d_k} = 1 - \dfrac{c}{|\inV(v_i)|} - c \cdot \tilde{\mu}$\;
\end{small}
\end{algorithm}

Based on the above observation, we propose an improved algorithm for computing $\tilde{d_k}$, as shown in Algorithm~\ref{alg:sample-small}. The algorithm first generates $n_r = O(\e_d^{-1} \log \delta_d^{-1})$ pairs of $\scw$-walks from randomly selected nodes in $I(v_k)$, and counts the number $cnt$ of pairs that meet (Lines 1-8). Then, it computes $\hat{\mu} = cnt/n_r$ as an estimation of $\mu$. If $\hat{\mu} \le \e_d$, then the algorithm determines that $n_r$ pairs of $\scw$-walks are sufficient for an accurate estimation of $\mu$; in that case, it terminates and returns an estimation of $d_k$ based on $\hat{\mu}$ (Lines 9-11).

On the other hand, if $\hat{\mu} > \e_d$, then the algorithm proceeds to generate a larger number of $\scw$-walks to derive a more accurate estimation of $\mu$. Towards this end, it first computes $\mu^* = \hat{\mu} + \sqrt{\hat{\mu} \cdot \e}$ as an upper bound of $\mu$, and uses $\mu^*$ to decide the total number $n_r^* = O(\mu^* \e_d^{-2} \log \delta_d^{-1})$ of $\scw$-walk pairs that are needed (Lines 12-13). After that, it increases the total number of $\scw$-walk pairs to $n_r^*$, and recounts the number $cnt$ of pairs that meet (Lines 14-19). Finally, it derives $\tilde{u}=cnt/n_r^*$ as an improved estimation of $\mu$, and returns an approximate correction factor $\tilde{d_k}$ computed based on $\tilde{\mu}$ (Lines 20-21).

The following lemmas establish the asymptotic guarantees of Algorithm~\ref{alg:sample-small}.
\begin{lemma} \label{lmm:opt-D-error}
With at least $1 - \delta_d$ probability, Algorithm~\ref{alg:sample-small} returns $\tilde{d_k}$ such that $|\tilde{d_k} - d| \le \e_d$ holds. \done
\end{lemma}

\begin{lemma} \label{lmm:opt-D-time}
Algorithm~\ref{alg:sample-small} generates $O(\frac{\mu+\e_d}{\e_d^2} \log \frac{1}{\delta_d})$ $\scw$-walks in expectation, and runs in $O(\frac{\mu+\e_d}{\e_d^2} \log \frac{1}{\delta_d})$ expected time. \done
\end{lemma}

By Lemma~\ref{lmm:opt-D-error}, Algorithm~\ref{alg:sample-small} uses a number of $\scw$-walks that is roughly $\max\{\mu, \e_d\}$ times the number in Algorithm~\ref{alg:naive-d}, which leads to significantly improved efficiency. In addition, we note that Algorithm~\ref{alg:sample-small} can be easily revised into a general method that estimates the expectation $\mu_z$ of a Bernoulli distribution by taking $O(\frac{\mu_z+\e}{\e^2} \log \frac{1}{\delta})$ samples, while ensuring at most $\e$ estimation error with at least $1-\delta$ success probability. In particular, the only major change needed is to replace each $\scw$-walk pair in Algorithm~\ref{alg:sample-small} with a sample from the Bernoulli distribution. In this context, we can prove that the number of samples used by Algorithm~\ref{alg:sample-small} is {\em asymptotically optimal}.

Specifically, let $z_1, z_2, \ldots$ be a sequence of i.i.d.\ Bernoulli random variables, and $\mu_Z = \E[z_i]$. Let $\mathcal{A}$ be an algorithm that inspects $z_i$ in ascending order of $i$, and stops at a certain $z_j$ before returning an estimation $\tilde{\mu_Z}$ of $\mu_Z$. In addition, for any possible sequence of $z_i$, $\mathcal{A}$ runs in finite expected time, and ensures that $|\tilde{\mu_Z}-\mu_Z| \le \e$ with at least $1-\delta$ probability. It can be verified that the revised Algorithm~\ref{alg:sample-small} is an instance of $\mathcal{A}$. The following lemma shows that no other instance of $\mathcal{A}$ can be asymptotically more efficient than Algorithm~\ref{alg:sample-small}.

\begin{lemma} \label{lmm:opt-D-optimal}
Any instance of $\mathcal{A}$ has $\Omega( \frac{\max\{\mu_z, \e\}}{\e^2} \log \frac{1}{\delta})$ expected time complexity when $\mu_z < 0.5$. \done
\end{lemma}

Our proof of Lemma~\ref{lmm:opt-D-optimal} utilizes an important result by Dagum et al.\ \cite{DKLR00} that establishes a lower bound of the expected time complexity of $\mathcal{A}$, when it provides a worst-case guarantee in terms of the relative error (instead of absolute error) in $\tilde{\mu_z}$. Dagum et al.\ \cite{DKLR00} also provide a sampling algorithm whose time complexity matches their lower bound, but the algorithm is inapplicable in our context, since it requires as input a relative error bound, which cannot be translated into an absolute error bound unless $\mu_z$ is known.

\subsection{Reduction of Space Consumption} \label{sec:opt-R}

Recall that our {\em SLING} index pre-computes a set $H(v_i)$ of approximate HPs for each node $v_i$, such that each $\tilde{h}^{(\l)}(v_i, v_k) \in H(v_i)$ is no smaller than a threshold $\theta = O(\e)$. The total size of all $H(v_i)$ is $O(n/\e)$, which is asymptotically near-optimal, but may still be costly from a practical perspective (especially when $\e$ is small). To address this issue, we aim to reduce the size of $H(v_i)$ without affecting the time complexity of {\em SLING}.

We observe that, in each $H(v_i)$, a significant portion of the approximate HPs are in the form of $\tilde{h}^{(1)}(v_i, v_k)$ or $\tilde{h}^{(2)}(v_i, v_k)$, i.e., they concern the HPs from $v_i$ to the nodes within two hops away from $v_i$. 
On the other hand, such HPs can be easily computed using a two-hop traversal from $v_i$, as we will show shortly. This leads to the following idea for space reduction: we remove from $H(v_i)$ all approximate HPs that are at steps $1$ and $2$, and we recompute those HPs {\em on the fly} during query processing. The re-computation may lead to slightly increased query cost, but as long as it takes $O(1/\e)$ time, it would not affect the asymptotic performance of {\em SLING}. In the following, we clarify how we implement this idea.

First, we present a simple and {\em precise} algorithm for computing the set $H^\prime(v_i)$ of HPs from node $v_i$ to other nodes at steps $1$ and $2$, as shown in Algorithm~\ref{alg:forward}. The algorithm first initializes a set $H^\prime(v_i) = \emptyset$ for storing HPs, and then inserts $h^{(0)}(v_i, v_i) = 1$ into $H^\prime(v)$. After that, for each in-neighbor $v_x$ of $v_i$, it sets $h^{(1)}(v_i, v_x) = \frac{\scw}{|I(v_i)|}$, which is the exact probability that a $\scw$-walk from $v_i$ would hit $v_x$ at step $1$. In turn, for each in-neighbor $v_y$ of $v_x$, the algorithm initializes $h^{(2)}(v_i, v_y) = \scw \cdot \frac{h^{(1)}(v_i, v_x)}{|I(v_x)|}$ in $H^\prime(v_i)$, if it is not yet inserted into $H^\prime(v_i)$; otherwise, the algorithm increases $h^{(2)}(v_i, v_y)$ by $\scw \cdot \frac{h^{(1)}(v_i, v_x)}{|I(v_x)|}$ in $H^\prime(v_i)$. This reason is that if a $\scw$-walk from $v_i$ hits $v_x$ at step $1$, then it has $\frac{\scw}{|I(v_x)|}$ probability to hit $v_y$ at step $2$. After all of $v_i$'s in-neighbors are processed, the algorithm terminates and returns $H^\prime(v_i)$.

\begin{algorithm}[t] \label{alg:forward}
\begin{small}
\caption{\em An algorithm for constructing $H^\prime(v_i)$}
\KwIn{a node $v_i$}
\KwOut {A set $H^\prime(v_i)$ of precise HPs from $v_i$ at steps $1$ and $2$}
\BlankLine
Initialize $H^\prime(v_i) = \emptyset$\;
Insert $h^{(0)}(v_i, v_i) = 1$ into $H^\prime(v)$\;
\For{each node $v_x \in I(v_i)$}
{
    Insert $h^{(1)}(v_i, v_x) = \frac{c}{|I(v_i)|}$ into $H^\prime(v)$\;
    \For{each node $v_y \in I(v_x)$}
    {
        \If{$h^{(2)}(v_i, v_y) \notin H^\prime(v_i)$}
        {
            Insert $h^{(2)}(v_i, v_y) = \scw \cdot \frac{h^{(1)}(v_i, v_x)}{|I(v_x)|}$ into $H^\prime(v_i)$\;
        }
        \Else
        {
            Increase $h^{(2)}(v_i, v_y)$ by $\scw \cdot \frac{h^{(1)}(v_i, v_x)}{|I(v_x)|}$ in $H^\prime(v_i)$\;
        }
    }
}
\Return $H^\prime(v_i)$
\end{small}
\end{algorithm}

Algorithm~\ref{alg:forward} runs in time linear to the total number $\eta(v_i)$ of incoming edges of $v_i$ and its in-neighbors, i.e.,
$$\eta(v_i) = |I(v_i)| + \sum_{v_x \in I(v_i)} |I(v_x)|.$$
If $\eta(v_i) = O(1/\e)$, then we can omit all step-$1$ and step-$2$ approximate HPs in $H(v_i)$, and compute them with Algorithm~\ref{alg:forward} during query processing without degrading the time complexity of {\em SLING}; otherwise, we need to retain all approximate HPs in $H(v_i)$. In our implementation of {\em SLING}, we set a constant $\gamma = 10$, and we exclude step-$1$ and step-$2$ HPs from $H(v_i)$ whenever $\eta(v_i) \le \gamma/\theta$, where $\theta = \Omega(\e)$ is the HP threshold used in the construction of $H(v_i)$ (see Algorithm~\ref{alg:backward}). Notice that each $\eta(v_i)$ can be computed in $O(|I(v_i)|)$ time by inspecting $v_i$ and all of its in-neighbors; therefore, the total computation cost of all $\eta(v_i)$ is $O(m)$, which does not affect {\em SLING}'s preprocessing time complexity. Furthermore, the on-the-fly computation of step-$1$ and step-$2$ HPs does not degrade {\em SLING}'s accuracy guarantee, since all HPs returned by Algorithm~\ref{alg:forward} are precise.

\subsection{Enhancement of Accuracy} \label{sec:opt-accuracy}

The approximation error of each $H(v_i)$ arises from the fact that it omits the HPs from $v_i$ that are smaller than a threshold $\theta$. A straightforward solution to reduce this error is to decrease $\theta$, but it would degrade the space overhead of $H(v_i)$. Instead, we propose to generate {\em additional} HPs in $H(v_i)$ {\em on-the-fly} during query processing, to increase the accuracy of query results.

Specifically, for each node $v_i$, after $H(v_i)$ is constructed (with the space reduction procedure in Section~\ref{sec:opt-R} applied), we inspect the set of approximate HPs $\tilde{h}^{(\l)}(v_i, v_j)$ in $H(v_i)$ such that $v_j$ has no more than $1/\sqrt{\e}$ in-neighbors, and then {\em mark} the $1/\sqrt{\e}$ largest HPs in the set. After that, whenever a SimRank query requires utilizing $H(v_i)$, we substitute $H(v_i)$ with an enhanced version $H^*(v_i)$ constructed on-the-fly. In particular, we first set $H^*(v_i) = H(v_i)$. Then, for every marked HP $\tilde{h}^{(\l)}(v_i, v_j)$ in $H(v_i)$, we process each in-neighbor $v_k$ of $v_j$ as follows:
\begin{itemize}[topsep = 6pt, parsep = 6pt, itemsep = 0pt, leftmargin=18pt]
\item If there exists $\tilde{h}^{(\l+1)}(v_i, v_k)$ in $H(v_i)$, then we omit $v_k$;

\item If $\tilde{h}^{(\l+1)}(v_i, v_k)$ is not in $H(v_i)$ and has not been inserted into $H^*(v_i)$, then we set $\displaystyle \tilde{h}^{(\l+1)}(v_i, v_k) = \frac{\sqrt{c}}{|I(v_j)|}h^{(\l)}(v_i, v_j)$, and insert it into $H^*(v_i)$;

\item Otherwise, we update $\tilde{h}^{(\l+1)}(v_i, v_k)$ in $H^*(v_i)$ as follows:
$\displaystyle \tilde{h}^{(\l+1)}(v_i, v_k) = \tilde{h}^{(\l+1)}(v_i, v_k) + \frac{\sqrt{c}}{|I(v_j)|}h^{(\l)}(v_i, v_j).$
\end{itemize}
In other words, if $H(v_i)$ does not contain an approximate HP from $v_i$ to $v_k$, then we generate $\tilde{h}^{(\l+1)}(v_i, v_k)$ in $H^*(v_i)$.

It can be verified that $0 < \tilde{h}^{(\l+1)}(v_i, v_k) \le h^{(\l+1)}(v_i, v_k)$, and hence, $H^*(v_i)$ provides higher accuracy than $H(v_i)$. In addition, the construction of $H^*(v_i)$ requires only $O(1/\e)$ time, and hence, it does not affect the $O(1/\e)$ query time complexity of {\em SLING}. Furthermore, marking HPs in all $H(v_i)$ requires only $O(n/\sqrt{\e})$ space and $O(n \log(1/\e) / \e)$ preprocessing time, which does not degrade the $O(n/\e)$ space and $\O\!\left(m/\e + n \log\frac{n}{\delta} / \e^2\right)$ preprocessing time complexity of {\em SLING}.

\subsection{Parallel and Out-of-Core Constructions} \label{sec:opt-parallel}

The preprocessing algorithms of {\em SLING} (i.e., Algorithms \ref{alg:naive-d}, \ref{alg:backward}, and \ref{alg:sample-small}) are {\em embarrassingly parallelizable}. In particular, Algorithm~\ref{alg:naive-d} (and Algorithm~\ref{alg:sample-small}) can be simultaneously applied to multiple nodes $v_k$ to compute the corresponding approximate correction factors $\tilde{d}_k$. Meanwhile, the main loop of Algorithm~\ref{alg:backward} (i.e., Lines 2-16) can be parallelized to construct the ``reverse'' HP sets $R_k$ for multiple nodes $v_k$ at the same time.

Furthermore, {\em SLING} does not require the complete index structure to fit in the main memory. Instead, we only need to keep all approximate correction factors $\tilde{v_k}$ ($k \in [1, n]$) in the memory, but can store the approximate HP set $H(v_x)$ for each node $v_x$ on the disk. To process a single-pair SimRank query on two nodes $v_i$ and $v_j$, we retrieve $H(v_i)$ and $H(v_j)$ from the disk and combine them with $\tilde{v_k}$ to derive the query result, which incurs a constant I/O cost, since $H(v_i)$ and $H(v_j)$ takes only $O(1/\e)$ space. In addition, the index construction process of {\em SLING} does not require maintaining all HP sets $H(v_x)$ simultaneously in the memory. Specifically, in Algorithm~\ref{alg:backward}, we can construct each ``reverse'' HP set $R_k$ in turn and write them to the disk; after that, we can construct all approximate HP sets $H(v_x)$ in a batch, by using an external sorting algorithm to sort all HPs $\tilde{h}^{(\l)}(v_x, v_k)$ by $v_x$. This process requires only $O(\frac{n}{\e}\log \frac{n}{\e})$ I/O accesses, since the total size of all $H(v_x)$ is $O(n/\e)$.

\section{Extension to Single-Source Queries} \label{sec:ext}

Given the {\em SLING} index introduced in Sections \ref{sec:our}, we can easily answer any single-source SimRank query from a node $v_i$, by invoking Algorithm~\ref{alg:single-pair} $n$ times to compute $s(v_i, v_j)$ for each node $v_j$. This leads to a total query cost of $O(n/\e)$, which is near-optimal since any single-source SimRank method requires $\Omega(n)$ time to output the results.
This straightforward algorithm, however, can be improved in terms of practical efficiency. To explain this, let us consider two nodes $v_i$ and $v_j$, such that $H(v_i)$ and $H(v_j)$ do not contain any HPs to the same node at the same step, i.e.,
$$\nexists v_k, \l, \;\; \tilde{h}^{(\l)}(v_i, v_k) \in H(v_i) \land \tilde{h}^{(\l)}(v_j, v_k) \in H(v_j).$$
Then, {\em SLING} would return $\tilde{s}(v_i, v_j) = 0$. We say that $H(v_i)$ and $H(v_j)$ do not {\em intersect} in this case. Intuitively, if we can avoid accessing those HP sets $H(v_j)$ that do not intersect with $H(v_i)$, then we can improve the efficiency of the single-source SimRank query from $v_i$. For this purpose, a straightforward approach is to maintain, for each combination of $v_k$ and $\l$, an {\em inverted list} $L(v_k, \l)$ that records the approximate HPs $\tilde{h}^{(\l)}(v_x, v_k)$ from any node $v_x$ to $v_k$. Then, to process a single-source SimRank query from node $v_i$, we first examine each approximate HP $\tilde{h}^{(\l)}(v_i, v_k) \in H(v_i)$ and retrieve $L(v_k, \l)$, based on which we compute $\tilde{s}(v_i, v_j)$ for any node $v_j$ with $\tilde{s}(v_i, v_j) > 0$.

Although the inverted list approach improves efficiency for single-source SimRank queries, it doubles the space consumption of {\em SLING}, since the inverted lists have the same total size as the approximate HP sets $H(v_i)$. Furthermore, the approach cannot be combined with the space reduction technique in Section~\ref{sec:opt-R}, because the former requires storing all approximate HPs in the inverted lists, whereas the latter aims to omit certain HPs to save space. To address this issue, we propose a single-source SimRank algorithm for {\em SLING} that finds a middle ground between the inverted list approach and the straightforward approach. The basic idea is that, given node $v_i$, we first retrieve all approximate HPs $\tilde{h}^{(\l)}(v_i, v_k) \in H(v_i)$, and then apply a variant of Algorithm~\ref{alg:backward} to compute the HPs from other nodes to each $v_k$; after that, we combine all HPs obtained to derive the query results. In other words, we construct the inverted lists relevant for the single-source query {\em on the fly}, instead of pre-computing them in advance.

\begin{algorithm}[t] \label{alg:single-source}
\begin{small}
\caption{\em An algorithm for single-source SimRank queries}
\KwIn {query node $v_i$ and threshold $\theta$}
\KwOut {an approximate SimRank score $\tilde{s}(v_i, v_j)$ for each node $v_j$}
\BlankLine
Initialize $\tilde{s}(v_i, v_j) = 0$ for all $v_j$\;
\For {each $\l$ such that $H(v_i)$ contains some approximate HP at step $\l$} {
    \For{each node $v_k$ such that $\tilde{h}^{(\l)}(v_i, v_k) \in H(v_i)$}
    {
        Initialize $\rho^{(0)}(v_k) = \tilde{h}^{(\l)}(v_i, v_k) \cdot d_k$\;
    }
    \For {$t = 1 $ to $\l$}
    {
        \For {each node $v_x$ such that $\rho^{(t-1)}(v_x) > (\sqrt{c})^{\l} \cdot \theta$}
        {
            \For {each out-neighbor $v_y$ of $v_x$}
            {
                \If{$\rho^{(t)}(v_y)$ does not exist}
                {
                    $\rho^{(t)}(v_y) = \frac{\scw}{|\inV(v_y)|} \cdot \rho^{(t-1)}(v_x)$\;
                }
                \Else
                {
                    $\rho^{(t)}(v_y) = \rho^{(t)}(v_y) + \frac{\scw}{|\inV(v_y)|} \cdot \rho^{(t-1)}(v_x)$\;
                }
            }
        }
    }
    \For{each $v_j$ such that $\rho^{(\l)}(v_j) > 0$}
    {
        $\tilde{s}(v_i, v_j) = \tilde{s}(v_i, v_j) + \rho^{(\l)}(v_j)$\;
    }
}
\Return $\tilde{s}(v_i, v_j)$ for each node $v_j$\;
\end{small}
\end{algorithm}

Algorithm~\ref{alg:single-source} shows the details of our method. It takes as input a query node $v_i$ and the threshold $\theta$ used in constructing $H(v_i)$ (see Algorithm~\ref{alg:backward}), and returns an approximate SimRank score $\tilde{s}(v_i, v_j)$ for each node $v_j$. The algorithm starts by initializing $\tilde{s}(v_i, v_j) = 0$ for all $v_j$ (Line 1). Then, it identifies the steps $\l$ such that there is at least one step-$\l$ approximate HP in $H(v_i)$; after that, it processes each of those steps in turn (Lines 2-10). The general idea of processing is as follows. By Equation~\ref{eqn:our-lsim}, if $v_i$ has a positive HP to a node $v_k$ at step $\l$, then for any other node $v_j$ with a positive HP to $v_k$ at step $\l$, we have $s(v_i, v_j) > 0$. To identify such nodes $v_j$ and their SimRank scores with $v_i$, we can apply the local update approach in Algorithm~\ref{alg:backward} to traverse $\l$ steps from $v_k$; however, the local update procedure needs to be slightly modified to deal with the fact that we may need to traverse from multiple $v_k$ simultaneously, i.e., when $v_i$ have positive HPs to multiple nodes at step $\l$.

Specifically, for each particular $\l$, Algorithm~\ref{alg:single-source} first identifies each node $v_k$ such that $\tilde{h}^{(\l)}(v_i, v_k) \in H(v_i)$, and initializes a temporary score $\rho^{(0)}(v_k) = \tilde{h}^{(\l)}(v_i, v_k)$ for $v_k$ (Line 3). After that, it traverses $\l$ steps from all $v_k$ simultaneously (Lines 5-8). In the $t$-th step ($t \in [1, \l]$), it inspects the temporary scores created in the $(t-1)$-th step, and omit those scores that are no larger than $(\scw)^\l \cdot \theta$ (Line 6). This omission is similar to the pruning of HPs applied in Algorithm~\ref{alg:backward}, except that the threshold used here is $(\scw)^\l$ times smaller than the threshold $\theta$ used in Algorithm~\ref{alg:backward}. The reason is that the local update procedure in Algorithm~\ref{alg:backward} starts from a node whose approximate HP equals $1$, whereas the procedure in Algorithm~\ref{alg:single-source} begins from a node whose temporary score $\rho^{(0)}(v_k) \le (\scw)^\l$, due to which we need to scale down the threshold to ensure accuracy.

For each temporary score $\rho^{(t-1)}(v_x)$ that is above the threshold, Algorithm~\ref{alg:backward} examines each out-neighbor $v_y$ of $v_x$, and checks whether the temporary score of $v_y$ at step $t$ (denoted as $\rho^{(t)}(v_y)$) exists. If it does not exist, then the algorithm initializes it as $\rho^{(t)}(v_y) = \frac{\scw}{|\inV(v_y)|} \cdot \rho^{(t-1)}(v_x)$; otherwise, the algorithm increases it by $\frac{\scw}{|\inV(v_y)|} \cdot \rho^{(t-1)}(v_x)$ (Lines 7-11). (Observe that this update rule is identical to that in Algorithm~\ref{alg:backward}.) Finally, after the $\l$-step traversal is finished, the algorithm adds each temporary score $\rho^{(\l)}(v_j)$ at step $\l$ into $\tilde{s}(v_i, v_j)$, and then proceeds to consider the next $\l$ (Lines 12-14). Once all steps $\l$ are processed, the algorithm returns each $\tilde{s}(v_i, v_j)$ as the final result.

We have the following lemma regarding the theoretical guarantees of Algorithm~\ref{alg:single-source}.
\begin{lemma} \label{lmm:ext-single-source}
Algorithm~\ref{alg:single-source} runs in $O\left(m \log^2\frac{1}{\e}\right)$ time, and ensures that each SimRank score returned has $\e$ worst-case error. \done
\end{lemma}

The time complexity of Algorithm~\ref{alg:single-source} is not as attractive as those of the inverted list approach and the straightforward approach, but is roughly comparable to the latter when $m = O(n/\e)$ (as is often the case in practice). In addition, we note that the time complexity of Algorithm~\ref{alg:single-source} matches that of the more recent method for single-source SimRank queries \cite{MKK14}, even though the latter relies on heuristic assumptions that do not hold in general (see Section~\ref{sec:exist-linear}).

\begin{figure*}[!t]
\begin{small}
\centering
\begin{tabular}{c}
\includegraphics[height=2.4mm]{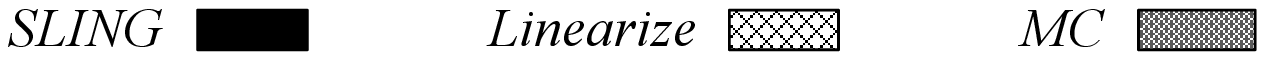} \vspace{-3mm}\\
\hspace{-2mm} \includegraphics[width=180mm]{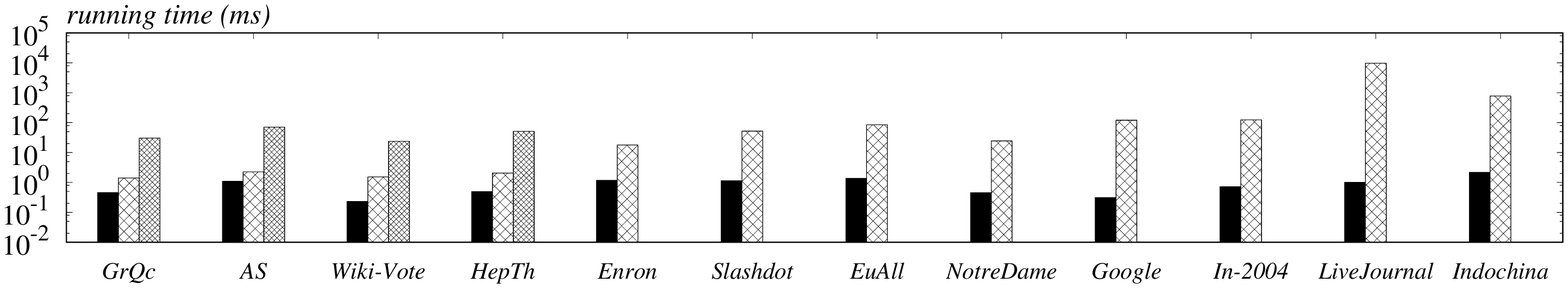} \vspace{0mm}
\end{tabular}
\figcapup  \vspace{-3mm}
\caption{Average query costs for single-pair SimRank queries.}
\label{fig:exp-pair}
\vspace{0mm}
\end{small}
\end{figure*}

\begin{figure*}[!t]
\begin{small}
\centering
\begin{tabular}{c}
\includegraphics[height=2.6mm]{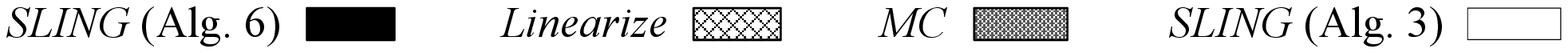} \vspace{-3mm}\\
\hspace{-2mm} \includegraphics[width=180mm]{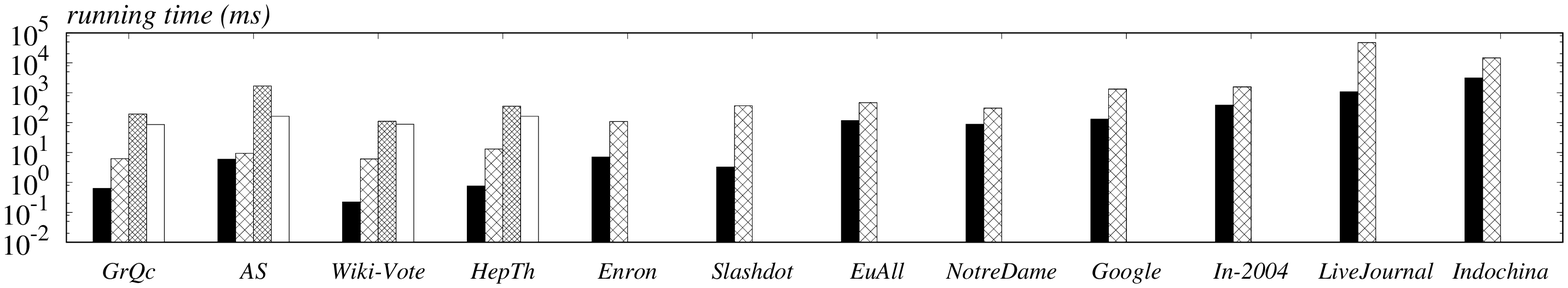} \vspace{0mm}
\end{tabular}
\figcapup  \vspace{-3mm}
\caption{Average query costs for single-source SimRank queries.}
\label{fig:exp-source}
\figcapdown
\vspace{-1mm}
\end{small}
\end{figure*}

\section{Experiments} \label{sec:exp}

\begin{table}[t]
\centering
\tblcapup
\caption{Datasets.}
\tblcapdown
\begin{small}
 \begin{tabular}{|l|l|r|r|} 
 \hline
 {\bf Dataset} & {\bf Type} & {\bf $\boldsymbol{n}$} & {\bf $\boldsymbol{m}$}	 \\ \hline
 GrQc	    &	undirected &	5,242	&	14,496 		\\
 AS	        &	undirected &	6,474	&	13,895 		\\
 Wiki-Vote	& 	directed &	7,155	&	103,689		\\
 HepTh	    & 	undirected &	9,877	&	25,998		\\
 Enron	    &	undirected &	36,692	&	183,831		\\
 Slashdot	&	directed &	77,360	&	905,468		\\
 EuAll	    &	directed &	265,214	&	400,045		\\
 NotreDame	&	directed &	325,728	&	1,497,134		\\
 Google	    &	directed &	875,713	&	5,105,049		\\
 In-2004	&	directed &	1,382,908	&	17,917,053		\\
 LiveJournal &	directed	&	4,847,571	&	68,993,773		\\
 Indochina	& 	directed &	7,414,866	&	194,109,311		\\
 \hline
\end{tabular}
\end{small}
\label{tbl:datasets}
\vspace{1mm}
\end{table}

This section experimentally evaluates {\em SLING}. Section~\ref{sec:exp-setting} clarifies the experimental settings, and Section~\ref{sec:exp-results} presents the experimental results.

\subsection{Experimental Settings} \label{sec:exp-setting}

\header
{\bf Datasets and Environment.} We use twelve graph datasets that are publicly available from \cite{SNAP,LWA} and are commonly used in the literature. Table~\ref{tbl:datasets} shows the statistics of each graph. We conduct all of experiments on a Linux machine with a 2.6GHz CPU and 64GB memory. All methods tested are implemented in C++. (Our code is available at \cite{SLING}.)

\header
{\bf Methods and Parameters.} We compare {\em SLING} against two state-of-the-art methods for SimRank computation: the linearization method \cite{MKK14,MKK15} (referred to as {\em Linearize}) and the Monte Carlo method \cite{FR05} (referred to as {\em MC}). {\em Linearize} has three parameters $T$, $R$, and $L$. Following the recommendations in \cite{MKK14}, we set $T = 11$, $R=100$, and $L=3$. In addition, we set the decay factor $c$ in the SimRank model to $0.6$, as suggested in previous work \cite{MKK14,Yu13,LVGT10,YuM15a,YuM15b}. Under this setting, {\em Linearize} ensures a worst-case error $\e = c^{T}/(1-c) \approx 0.01$ in each SimRank score, {\em if it is able to derive an exact diagonal correction matrix $D$}. However, as we discuss in Section~\ref{sec:exist-linear}, {\em Linearize} utilizes an approximate version of $D$ that provides no quality assurance, due to which the above error bound does not hold.

For {\em SLING}, we set its maximum error $\e = 0.025$, which is roughly comparable to the quality assurance of the linearization method given a precise $D$. Towards this end, we set $\e_d = 0.005$ and $\theta = 0.000725$, which ensures $\e < 0.025$ by Theorem~\ref{thm:err-bound}. In addition, we set $\delta_d = 1/n^2$, which guarantees that the preprocessing algorithm of {\em SLING} succeeds with at least $1 - 1/n$ probability. For {\em MC}, we set $\e = 0.025$, as in {\em SLING}.

\subsection{Experimental Results} \label{sec:exp-results}

In the first set of experiments, we randomly generate $1000$ single-pair SimRank queries on each dataset, and evaluate the average computation time of each method in answering the queries. Figure~\ref{fig:exp-pair} shows the results. We omit {\em MC} on all but the four smallest datasets, since its index size exceeds 64GB on the large graphs. Observe that the query time of {\em SLING} is at most $2.2$ms in all cases, and is often several orders of magnitude smaller than that of {\em Linearize}. In particular, on {\em LiveJournal}, {\em SLING} is around $10000$ times faster than {\em Linearize}. This is consistent with the fact that {\em SLING} and {\em Linearize} has $O(1/\e)$ and $O(m \log\frac{1}{\e})$ query time complexities, respectively. Meanwhile, {\em Linearize} incurs a smaller query cost than {\em MC} on the four smallest datasets, which is also observed in previous work \cite{MKK14}.

Our second set of experiments evaluates the average computation cost of each method in answering $500$ random single-source SimRank queries. For {\em SLING}, we consider two different methods: one that directly uses Algorithm~\ref{alg:single-source}, and another one that invokes Algorithm~\ref{alg:single-pair} once for each node. Figure~\ref{fig:exp-source} illustrates the results. Notice that the method that applies Algorithm~\ref{alg:single-pair} is significantly slower than Algorithm~\ref{alg:single-source}, even though the former (resp.\ latter) runs in $O(n/\e)$ time (resp.\ $O(m \log^2\frac{1}{\e})$ time). This is in accordance with our analysis in Section~\ref{sec:ext}, which shows that adopting Algorithm~\ref{alg:single-pair} for single-source queries would incur unnecessary overheads and lead to inferior query time. Since the method that employs Algorithm~\ref{alg:single-pair} is not competitive, we omit it on all but the four smallest datasets.

\begin{figure*}[!t]
\begin{small}
\centering
\begin{tabular}{c}
\includegraphics[height=2.4mm]{./figs/legend-single.eps} \vspace{-3mm}\\
\hspace{-2mm} \includegraphics[width=180mm]{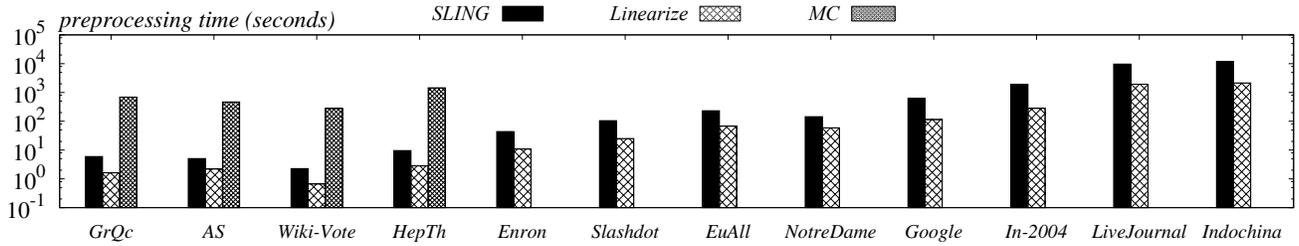} \vspace{0mm}
\end{tabular}
\figcapup  \vspace{-2.5mm}
\caption{Preprocessing cost of each method.}
\label{fig:exp-precompute}
\vspace{1mm}
\end{small}
\end{figure*}

\begin{figure*}[!t]
\begin{small}
\centering
\begin{tabular}{c}
\includegraphics[height=2.4mm]{./figs/legend-single.eps} \vspace{-3mm}\\
\hspace{-2mm} \includegraphics[width=180mm]{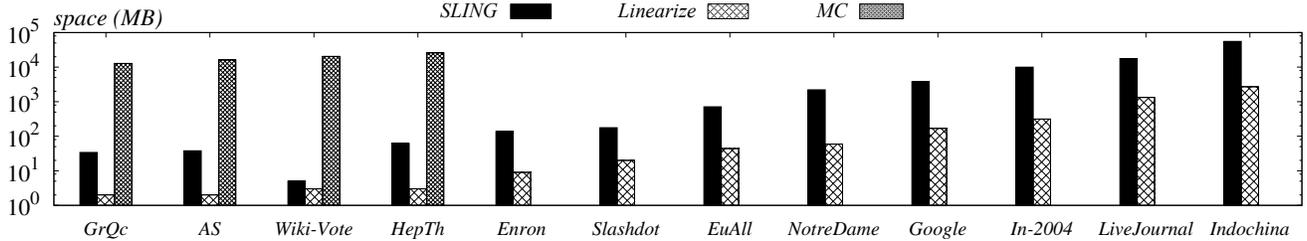} \vspace{0mm}
\end{tabular}
\figcapup  \vspace{-2.5mm}
\caption{Space consumption of each method.}
\label{fig:exp-space}
\vspace{1mm}
\end{small}
\end{figure*}

\begin{figure*}[!t]
\begin{small}
\centering
\begin{tabular}{cccc}
\multicolumn{4}{c}{\hspace{-4mm} \includegraphics[height=3.5mm]{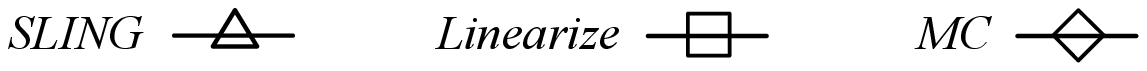}} \vspace{-0.5mm} \\
\hspace{-6mm}\includegraphics[height=35mm]{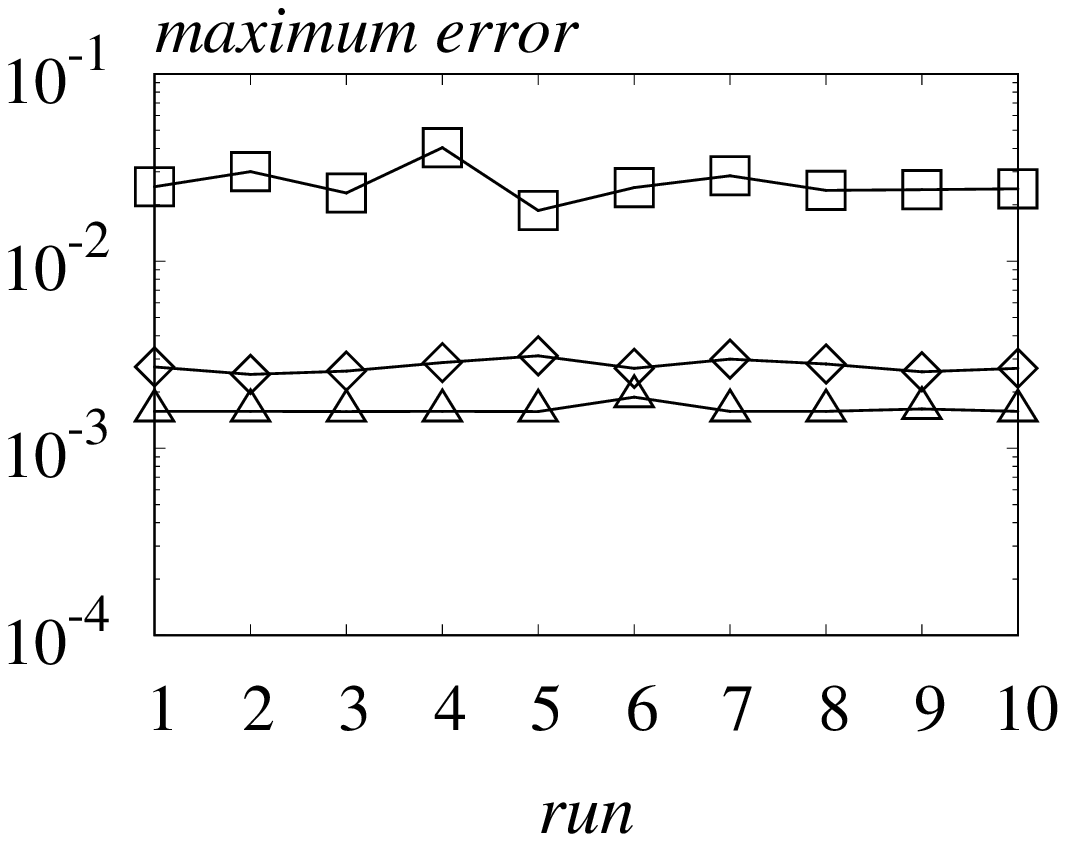}
&
\hspace{-8.5mm}\includegraphics[height=35mm]{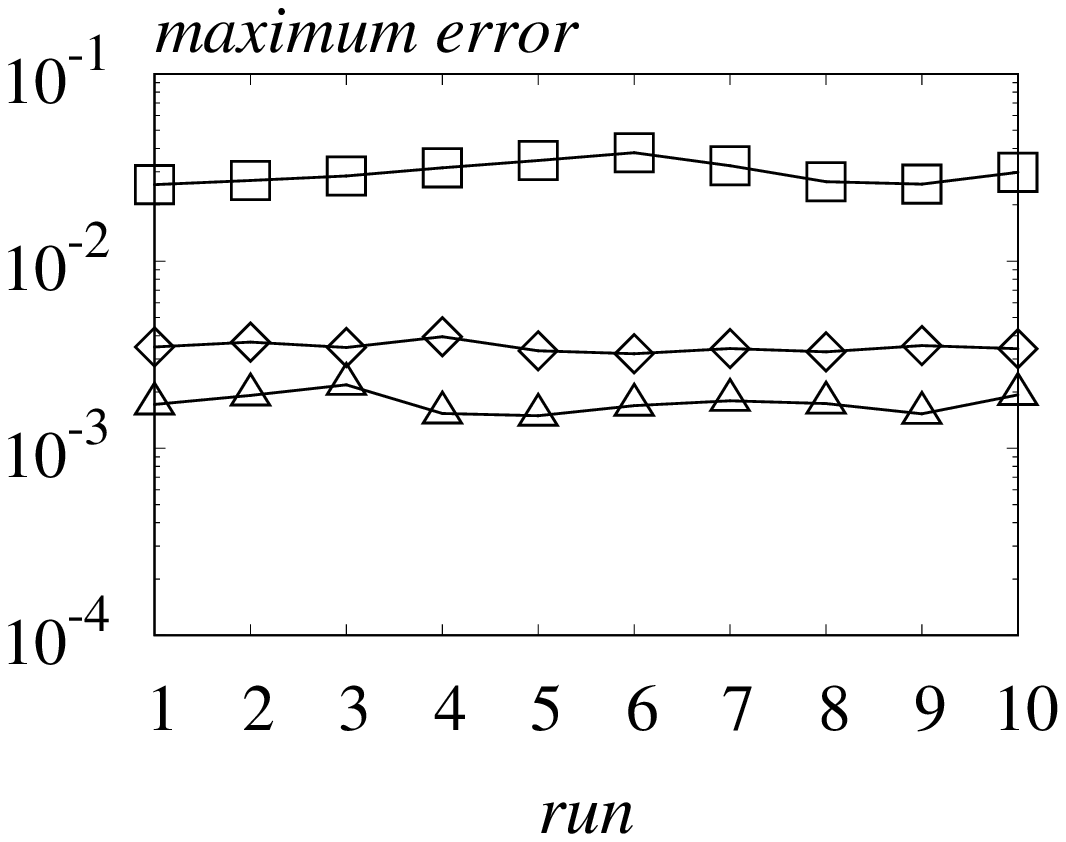}
&
\hspace{-8.5mm}\includegraphics[height=35mm]{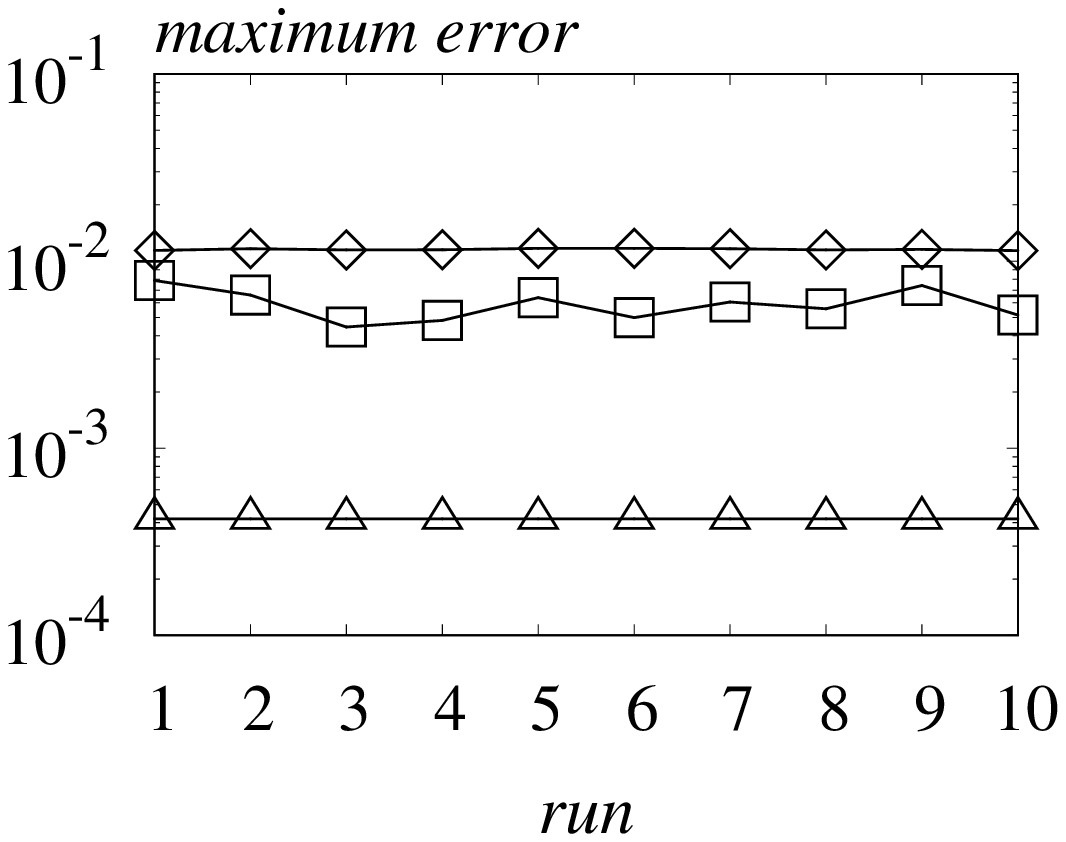}
&
\hspace{-8.5mm}\includegraphics[height=35mm]{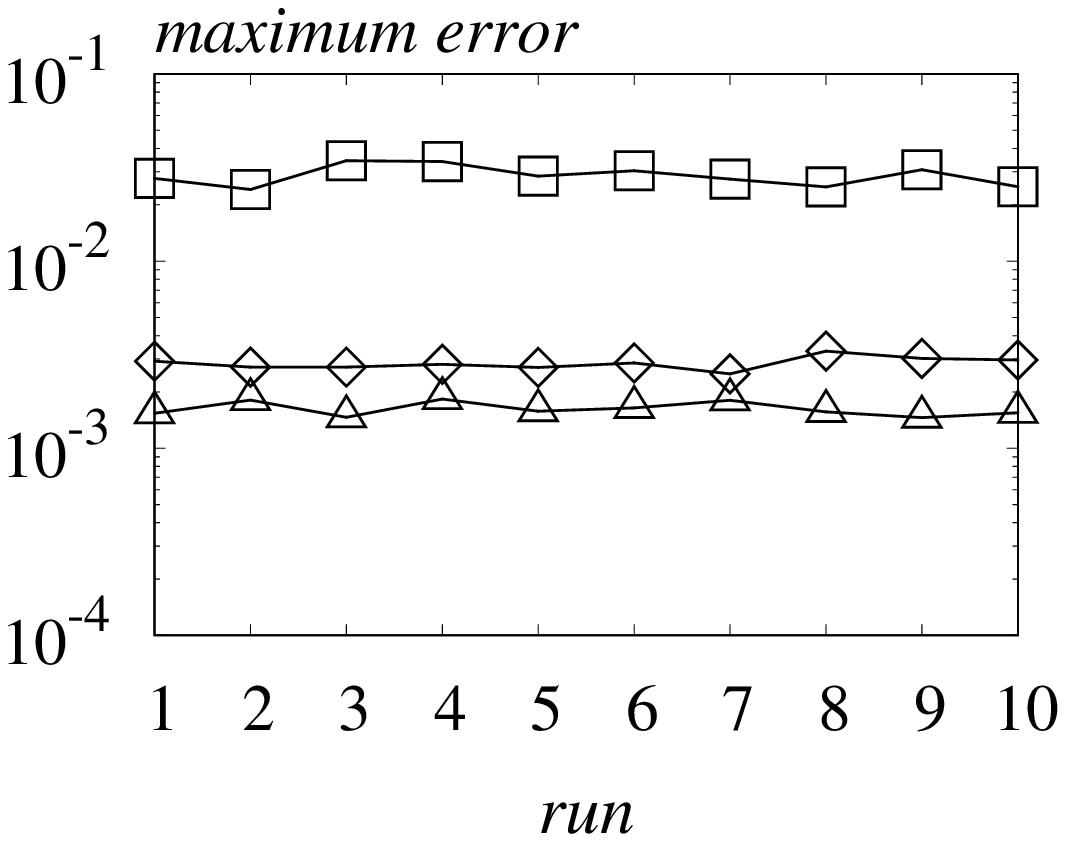}
\\
\hspace{-4mm} (a)  GrQc & \hspace{-6mm}(b) AS
& \hspace{-6mm}(c) Wiki-Vote & \hspace{-6mm}(d) HepTh
\end{tabular}
\figcapup  \vspace{-1mm} \caption{Maximum SimRank error of each method measured in $\boldsymbol{10}$ different runs.} 
\label{fig:worst-error}
\vspace{-2mm}
\end{small}
\end{figure*}

Among all methods for single-source SimRank queries, {\em SLING} (with Algorithm~\ref{alg:single-source}) achieves the best performance, but its improvement over {\em Linearize} is less pronounced when compared with the case of single-pair queries. This, as we mention in Section~\ref{sec:ext}, is because the local update procedure in Algorithm~\ref{alg:single-source} incurs super-linear overheads, due to which the algorihtm's time complexity is the same as {\em Linearize}'s. Nonetheless, {\em SLING} is still at least $9$ times faster than {\em Linearize} on $7$ out of the $12$ datasets, and is $110$ times more efficient on {\em Slashdot}. Meanwhile, {\em MC} is consistently outperformed by {\em Linearize}.

Next, we plot the the preprocessing cost (resp.\ space consumption) of each method in Figure~\ref{fig:exp-precompute} (resp.\ Figure~\ref{fig:exp-space}). {\em Linearize} incurs a smaller pre-computation cost than {\em SLING} does; in turn, {\em SLING} is more efficient than {\em MC} in terms of pre-computation. The index size of {\em SLING} is considerably larger than {\em Linearize}, since {\em SLING} has an $O(n/\e)$ space complexity, while {\em Linearize} only incurs $O(n+m)$ space overhead. Nevertheless, {\em SLING} outperforms {\em MC} in terms of space efficiency. Overall, {\em SLING} is inferior to {\em Linearize} in terms of space overheads and preprocessing costs, but this is justified by the fact that {\em SLING} offers superior query efficiency and rigorous accuracy guarantee, whereas {\em Linearize} incurs significantly larger query costs and does not offer non-trivial bounds on its query errors. Furthermore, the pre-computation algorithm of {\em SLING} can be easily parallelized, as we discuss in Section~\ref{sec:opt-parallel} and demonstrate in Appendix~\ref{apnx:experiment}.

Our last three experiments focus on the query accuracy of each method. We first apply the power method (see Section~\ref{sec:exist-power}) on each of the four smallest graphs to compute the SimRank score of each node pair, setting the number of iterations in the method to $50$ (which results in a worst-case error below $10^{-11}$). We take the SimRank scores thus obtained as the ground truth, and use them to gauge the error of each method computing all-pair SimRank scores. We do not repeat this experiment on larger graphs, due to the tremendous overheads in computing all-pair SimRank results.

Figure~\ref{fig:worst-error} illustrates the maximum query error incurred by each method in all-pair SimRank computation over $10$ different runs, where each run rebuilds the index of each method from scratch. Observe that the maximum error of {\em SLING} is always below $0.0025$, which is considerably smaller than the stipulated error bound $\e = 0.025$. {\em MC}'s maximum error is also below $\e = 0.025$, but is consistently larger than that of {\em SLING}, and is over $0.01$ on {\em Wiki-Vote}. In contrast, the maximum error of {\em Linearize} is above $0.025$ in most runs on {\em GrQc}, {\em AS}, and {\em Hepth}, which is consistent with our analysis that {\em Linearize} does not offer any worst-case guarantee in terms of query accuracy.

\begin{figure*}[!t]
\begin{small}
\centering
\begin{tabular}{cccc}
\multicolumn{4}{c}{\hspace{-4mm} \includegraphics[height=3.5mm]{./figs_revision/legend/legend2.eps}} \vspace{-0.5mm} \\
\hspace{-6mm}\includegraphics[height=35mm]{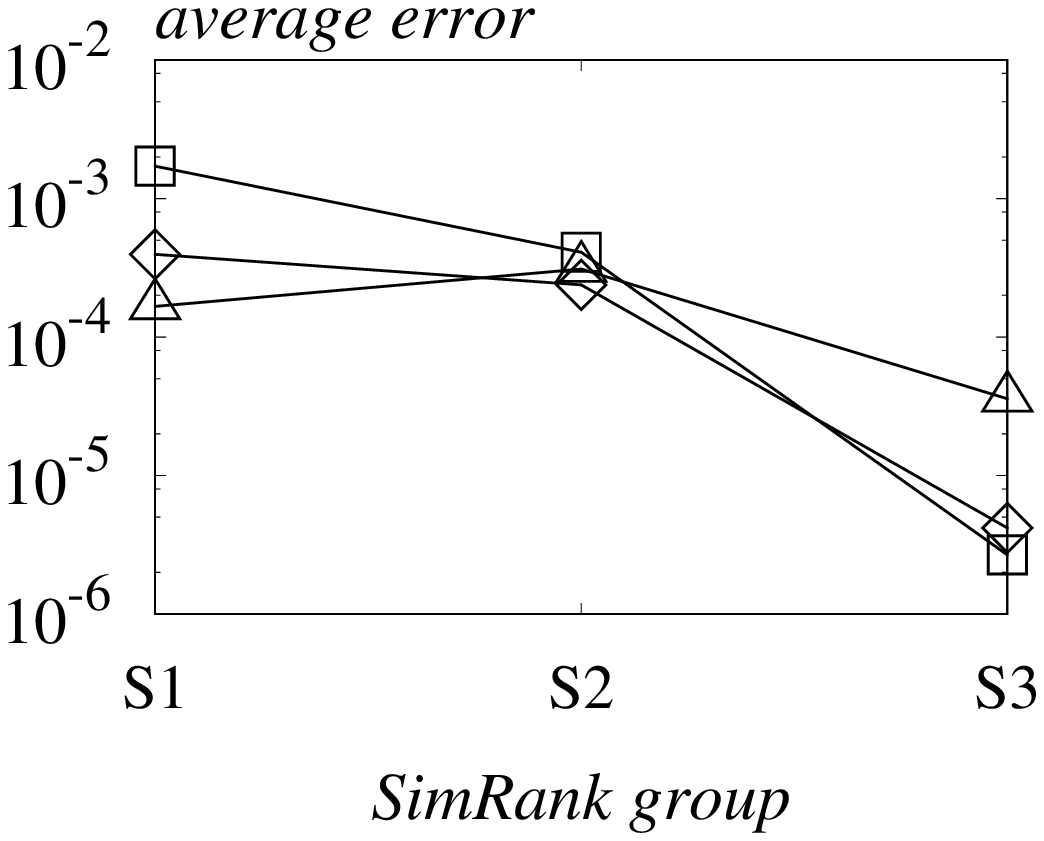}
&
\hspace{-8.5mm}\includegraphics[height=35mm]{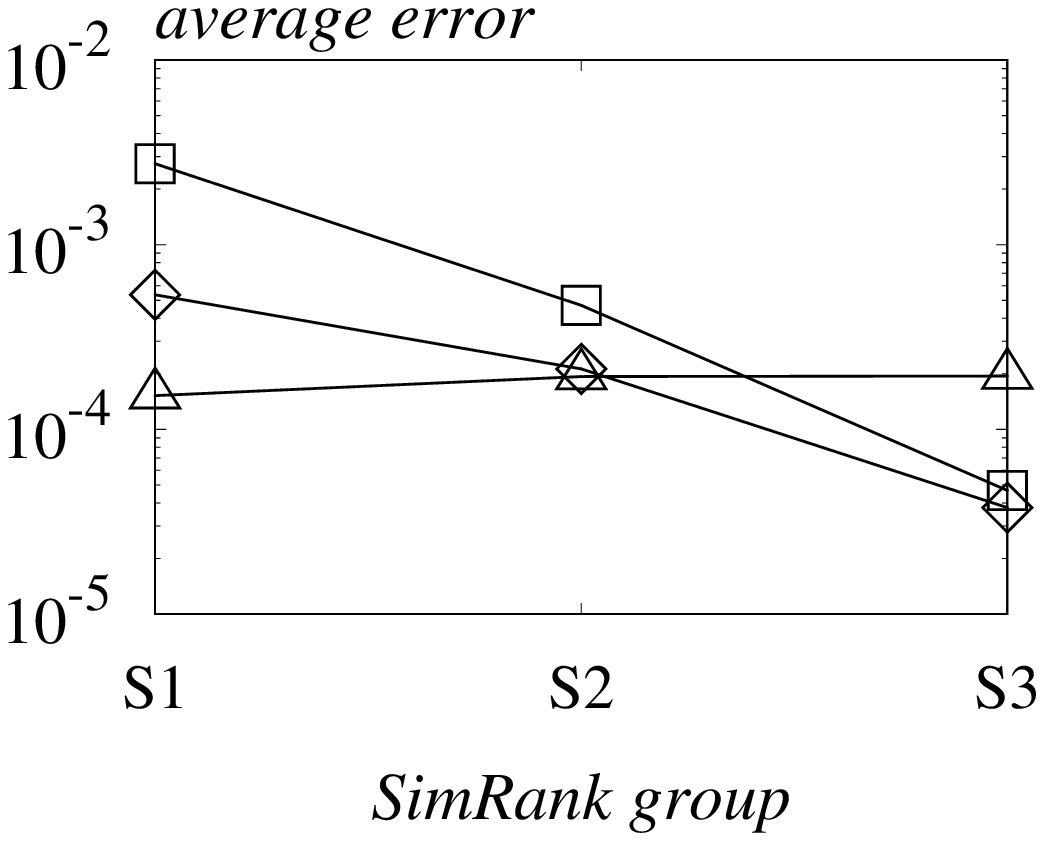}
&
\hspace{-8.5mm}\includegraphics[height=35mm]{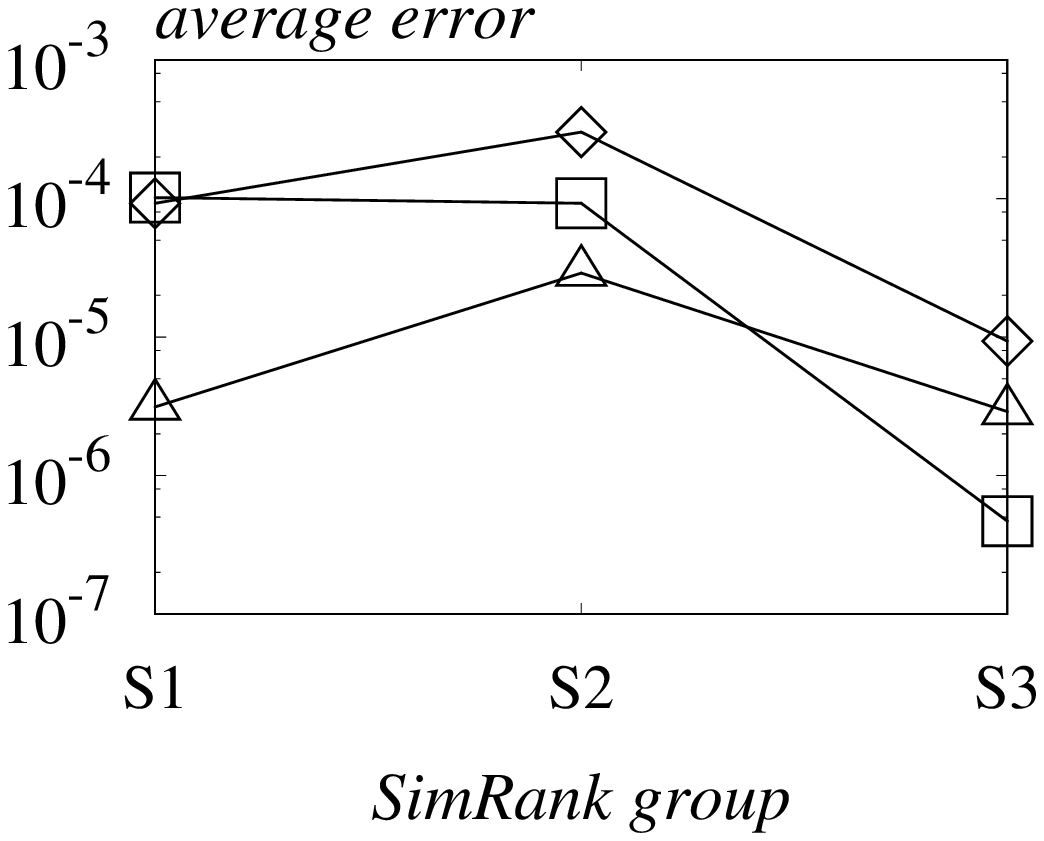}
&
\hspace{-8.5mm}\includegraphics[height=35mm]{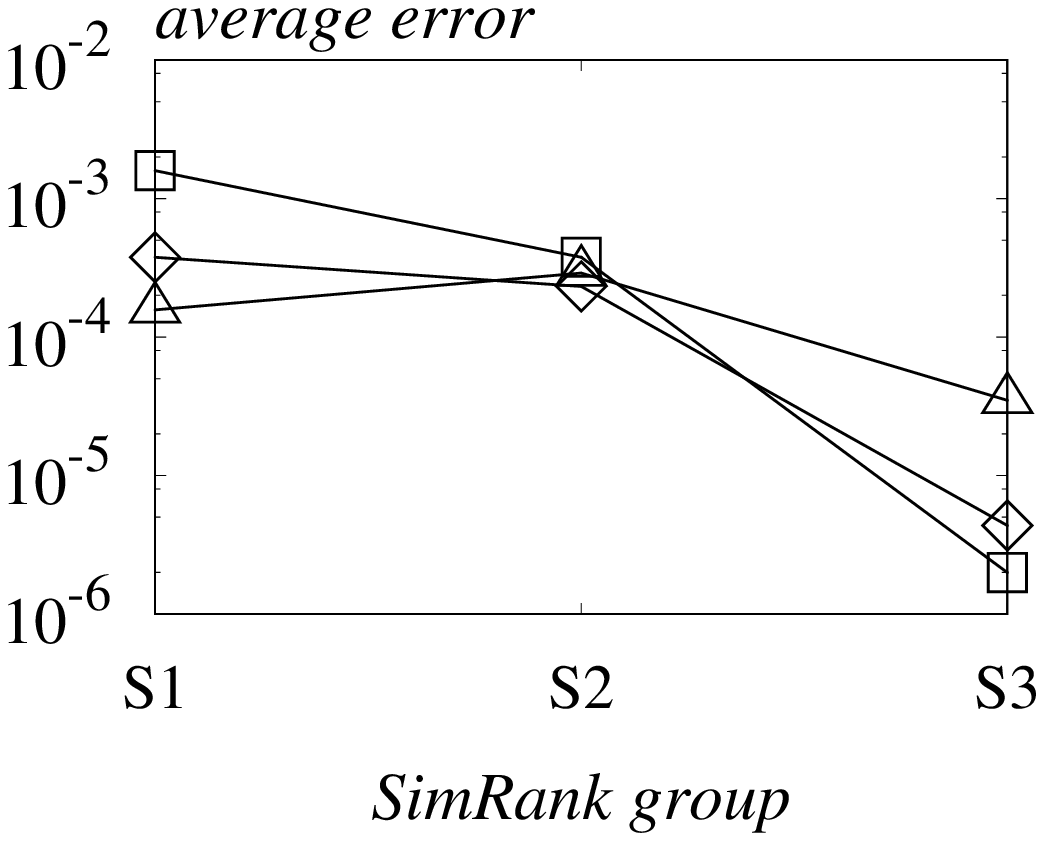}
\\
\hspace{-4mm} (a)  GrQc & \hspace{-6mm}(b) AS
& \hspace{-6mm}(c) Wiki-Vote & \hspace{-6mm}(d) HepTh
\end{tabular}
\figcapup \vspace{-1mm} \caption{Average SimRank error vs. SimRank group.} 
\label{fig:avg-error}
\end{small}
\end{figure*}

\begin{figure*}[!t]
\begin{small}
\centering
\begin{tabular}{cccc}
\multicolumn{4}{c}{\hspace{-4mm} \includegraphics[height=2.4mm]{./figs/legend-single.eps}} \vspace{0.5mm} \\
\hspace{-1mm}\includegraphics[height=32mm]{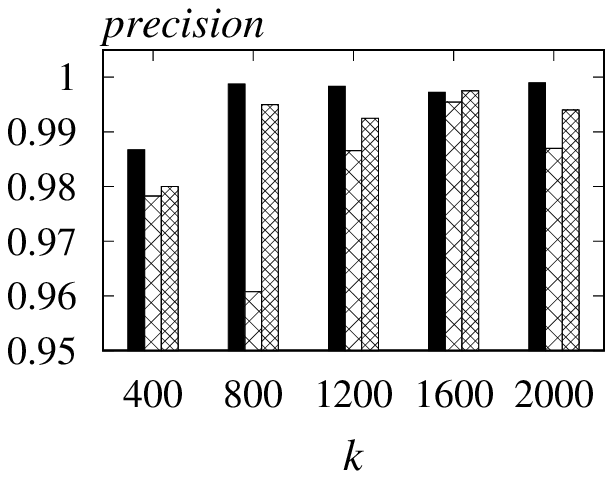}
&
\hspace{-0mm}\includegraphics[height=32mm]{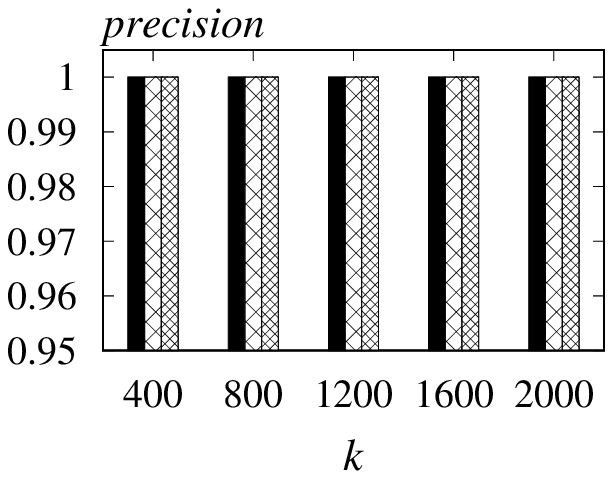}
&
\hspace{-0mm}\includegraphics[height=32mm]{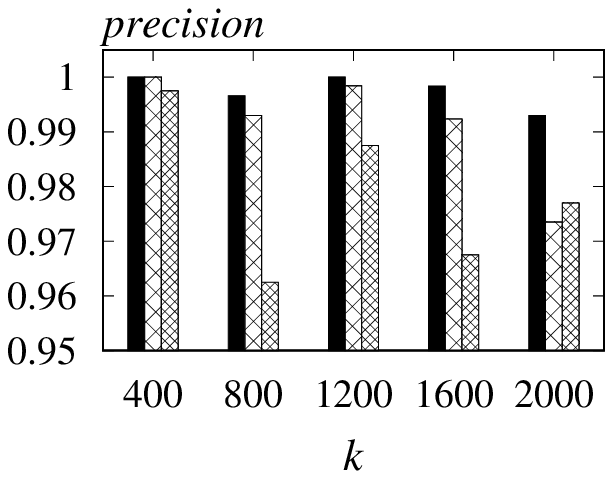}
&
\hspace{-0mm}\includegraphics[height=32mm]{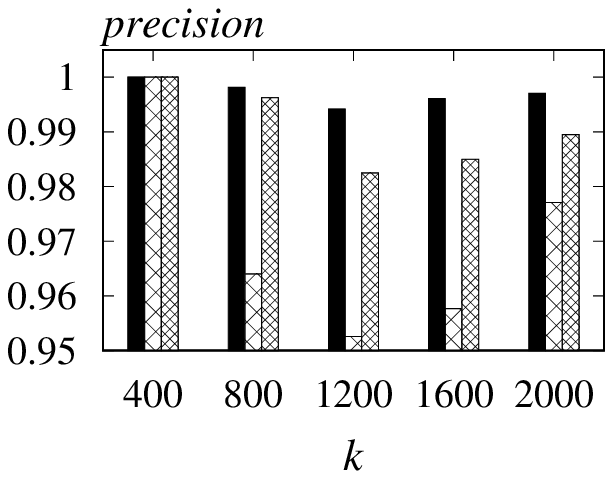}
\\
\hspace{2mm} (a)  GrQc & \hspace{2mm}(b) AS
& \hspace{2mm}(c) Wiki-Vote & \hspace{2mm}(d) HepTh
\end{tabular}
\figcapup  \vspace{-1mm} \caption{Precision of the top-$\boldsymbol{k}$ SimRank pairs returned by each method.} 
\label{fig:top-k-pair}
\vspace{-2mm}
\end{small}
\end{figure*}

To further assess each method's query accuracy, we divide the ground-truth SimRank scores into three groups $S_1$, $S_2$, and $S_3$, such that $S_1$ (resp.\ $S_2$) contains SimRank scores in the range of $[0.1, 1]$ (resp.\ $[0.01, 0.1]$), while $S_3$ concerns SimRank scores smaller than $0.01$. Intuitively, the scores in $S_1$ and $S_2$ are more important than those in $S_3$, since the former correspond to node pairs that are highly similar. Figure~\ref{fig:avg-error} shows the average query errors of each method for $S_1$, $S_2$, and $S_3$. Observe that, compared with {\em Linearize}, {\em SLING} incurs much smaller (resp.\ slightly smaller) errors on $S_1$ (resp.\ $S_2$). This indicates that {\em SLING} is more effective than {\em Linearize} in measuring the similarity of important node pairs. Meanwhile, {\em MC} is less accurate than {\em SLING} on $S_1$, and is considerably outperformed by both {\em SLING} and {\em Linearize} on {\em Wiki-Vote}.

Finally, we use the all-pair SimRank scores computed by each method to identify the $k$ node pairs with the highest SimRank scores\footnote{Note that we ignore any node pair containing two nodes that are identical.}, and we measure the {\em precision} of those $k$ pairs, i.e., the fraction of them among the ground-truth top-$k$ pairs. Figure~\ref{fig:top-k-pair} illustrates the results when $k$ varies from $400$ to $2000$. The precision of {\em SLING} is never worse than that of {\em Linearize}, and is up to $4\%$ higher than the latter in many cases. This is consistent with our results in Figure~\ref{fig:avg-error} that, for node pairs with large SimRank scores, {\em SLING} provides much higher accuracy than {\em Linearize} does. Meanwhile, {\em MC} yields lower accuracy than {\em SLING} does, and is significantly outperformed by both {\em SLING} and {\em Linearize} on {\em Wiki-Vote}. These results are also in agreement with those in Figure~\ref{fig:avg-error}.

\section{Other Related Work} \label{sec:related}

The previous sections have discussed the existing techniques that are most relevant to ours. In what follows, we survey other related work on SimRank computation. First, there is a line of research \cite{FNSO13,He10,Yu13,Li10,Yu14,YuM15b} on SimRank queries based on the following formulation of SimRank:
$$S = cP^{\top}SP + (1-c)I,$$
where $S$, $P$, and $T$ are $n \times n$ matrices such that $S(i, j) = s(v_i, v_j)$ for any $i, j$, $P$ is as defined in Equation~\ref{eqn:exist-linear-transit}, and $I$ is an identity matrix. However, as point out by Kusumoto et al.\ \cite{KMK14}, the above formulation is {\em incorrect} since it assumes that $(1-c)I$ equals the diagonal correction matrix $D$ (see Equation~\ref{eqn:def-lsim}), which does not hold in general. As a consequence, the methods in \cite{FNSO13,He10,Yu13,Li10,Yu14,YuM15b} fail to offer any guarantees in terms of the accuracy of SimRank scores, due to which we do not consider them in this paper.

Second, several variants \cite{AMC08,FR05,Lin12,YuM15a,ZhaoHS09} of SimRank have been proposed to enhance the quality of similarity measure and mitigate certain limitations of SimRank. Antonellis et al.\ \cite{AMC08} present {\em SimRank++}, which extends SimRank by taking into account the weights of edges and prior knowledge of node similarities. Jin et al.\ \cite{Jin11} introduce {\em RoleSim}, which guarantees to recognize automorphically or structurally equivalent nodes. Fogaras and R{\'{a}}cz \cite{FR05} propose {\em PSimRank}, which improves the quality of SimRank by allowing random walks that are close to each other to have a higher probability to meet. Yu and McCann \cite{YuM15a} present {\em SimRank${}^{\#}$}, which defines the similarity between two nodes based on the {\em consine similarity} of their neighbors. Zhao et al.\ \cite{ZhaoHS09} introduce {\em P-Rank}, which consider both in-neighbors and out-neighbors of two nodes when measuring their similarity.

Finally, there is existing work \cite{LeeLY12,KMK14,FNSO13,TaoYL14,MKK15,ZhengZF0Z13} that studies {\em top-$k$ SimRank queries} and {\em SimRank similarity joins}. In particular, a top-$k$ SimRank queries takes as input a node $v_i$, and asks for the $k$ nodes $v_j$ with the largest SimRank score $s(v_i, v_j)$. Meanwhile, a SimRank similarity join asks for all pairs of nodes whose SimRank scores are among the largest $k$, or are larger than a predefined threshold. Techniques designed for these two types of queries are generally inapplicable for single-pair and single-source SimRank queries.

\section{Conclusions} \label{sec:conclusions}

This paper presents the {\em SLING} index for answering single-pair and single-source SimRank queries with $\e$ worst-case error in each SimRank score. {\em SLING} requires $O(n/\e)$ space and ${O(m/\e + n\log \frac{n}{\delta}/\e^2)}$ pre-computation time, and it handles any single-pair (resp.\ single-source) query in $O(1/\e)$ (resp.\ $O(n/\e)$) time. The space and query time complexities of {\em SLING} are near-optimal, and are significantly better than those of the existing solutions. In addition, {\em SLING} incorporates several optimization techniques that considerably improves its practical performance. Our experiments show that {\em SLING} provides superior query efficiency against the states of the art. For future work, we plan to (i) investigate techniques to reduce the index size of {\em SLING}, and (ii) extend {\em SLING} to handle other similarity measures for graphs.

\allowdisplaybreaks

\begin{small}
\bibliographystyle{abbrv}
\bibliography{ref}

\begin{thebibliography}{10}

\bibitem{SNAP}
\url{http://snap.stanford.edu/data/index.html}.

\bibitem{LWA}
\url{http://law.di.unimi.it/datasets.php}.

\bibitem{SLING}
\url{https://sourceforge.net/projects/slingsimrank/}.

\bibitem{ACL06}
R.~Andersen, F.~R.~K. Chung, and K.~J. Lang.
\newblock Local graph partitioning using pagerank vectors.
\newblock In {\em {FOCS}}, pages 475--486, 2006.

\bibitem{AMC08}
I.~Antonellis, H.~G. Molina, and C.~C. Chang.
\newblock Simrank++: query rewriting through link analysis of the click graph.
\newblock {\em PVLDB}, 1(1):408--421, 2008.

\bibitem{ChungL06}
F.~R.~K. Chung and L.~Lu.
\newblock Concentration inequalities and martingale inequalities: {A} survey.
\newblock {\em Internet Mathematics}, 3(1):79--127, 2006.

\bibitem{DKLR00}
P.~Dagum, R.~M. Karp, M.~Luby, and S.~M. Ross.
\newblock An optimal algorithm for monte carlo estimation.
\newblock {\em {SIAM} J. Comput.}, 29(5):1484--1496, 2000.

\bibitem{FR05}
D.~Fogaras and B.~R{\'{a}}cz.
\newblock Scaling link-based similarity search.
\newblock In {\em {WWW}}, pages 641--650, 2005.

\bibitem{FRCS05}
D.~Fogaras, B.~R{\'{a}}cz, K.~Csalog{\'{a}}ny, and T.~Sarl{\'{o}}s.
\newblock Towards scaling fully personalized pagerank: Algorithms, lower
  bounds, and experiments.
\newblock {\em Internet Mathematics}, 2(3):333--358, 2005.

\bibitem{FNSO13}
Y.~Fujiwara, M.~Nakatsuji, H.~Shiokawa, and M.~Onizuka.
\newblock Efficient search algorithm for simrank.
\newblock In {\em ICDE}, pages 589--600, 2013.

\bibitem{GV12}
G.~H. Golub and C.~F. Van~Loan.
\newblock {\em Matrix Computations}.
\newblock Johns Hopkins University Press, 3 edition, 2012.

\bibitem{GGL13}
P.~Gupta, A.~Goel, J.~Lin, A.~Sharma, D.~Wang, and R.~Zadeh.
\newblock {WTF:} the who to follow service at twitter.
\newblock In {\em {WWW}}, pages 505--514, 2013.

\bibitem{He10}
G.~He, H.~Feng, C.~Li, and H.~Chen.
\newblock Parallel simrank computation on large graphs with iterative
  aggregation.
\newblock In {\em KDD}, pages 543--552, 2010.

\bibitem{JW02}
G.~Jeh and J.~Widom.
\newblock Simrank: a measure of structural-context similarity.
\newblock In {\em {SIGKDD}}, pages 538--543, 2002.

\bibitem{JehW03}
G.~Jeh and J.~Widom.
\newblock Scaling personalized web search.
\newblock In {\em {WWW}}, pages 271--279, 2003.

\bibitem{Jin11}
R.~Jin, V.~E. Lee, and H.~Hong.
\newblock Axiomatic ranking of network role similarity.
\newblock In {\em KDD}, pages 922--930, 2011.

\bibitem{KMK14}
M.~Kusumoto, T.~Maehara, and K.~Kawarabayashi.
\newblock Scalable similarity search for simrank.
\newblock In {\em {SIGMOD}}, pages 325--336, 2014.

\bibitem{LeeLY12}
P.~Lee, L.~V.~S. Lakshmanan, and J.~X. Yu.
\newblock On top-k structural similarity search.
\newblock In {\em {ICDE}}, pages 774--785, 2012.

\bibitem{Li10}
C.~Li, J.~Han, G.~He, X.~Jin, Y.~Sun, Y.~Yu, and T.~Wu.
\newblock Fast computation of simrank for static and dynamic information
  networks.
\newblock In {\em EDBT}, pages 465--476, 2010.

\bibitem{LLYHD10}
P.~Li, H.~Liu, J.~X. Yu, J.~He, and X.~Du.
\newblock Fast single-pair simrank computation.
\newblock In {\em {SDM}}, pages 571--582, 2010.

\bibitem{NK07}
D.~Liben{-}Nowell and J.~M. Kleinberg.
\newblock The link-prediction problem for social networks.
\newblock {\em {JASIST}}, 58(7):1019--1031, 2007.

\bibitem{Lin12}
Z.~Lin, M.~R. Lyu, and I.~King.
\newblock Matchsim: a novel similarity measure based on maximum neighborhood
  matching.
\newblock {\em KAIS}, 32(1):141--166, 2012.

\bibitem{LVGT10}
D.~Lizorkin, P.~Velikhov, M.~N. Grinev, and D.~Turdakov.
\newblock Accuracy estimate and optimization techniques for simrank
  computation.
\newblock {\em {VLDB} J.}, 19(1):45--66, 2010.

\bibitem{MKK14}
T.~Maehara, M.~Kusumoto, and K.~Kawarabayashi.
\newblock Efficient simrank computation via linearization.
\newblock {\em CoRR}, abs/1411.7228, 2014.

\bibitem{MKK15}
T.~Maehara, M.~Kusumoto, and K.~Kawarabayashi.
\newblock Scalable simrank join algorithm.
\newblock In {\em {ICDE}}, pages 603--614, 2015.

\bibitem{RotheS14}
S.~Rothe and H.~Sch{\"{u}}tze.
\newblock Cosimrank: {A} flexible {\&} efficient graph-theoretic similarity
  measure.
\newblock In {\em {ACL}}, pages 1392--1402, 2014.

\bibitem{SH11}
N.~Spirin and J.~Han.
\newblock Survey on web spam detection: principles and algorithms.
\newblock {\em {SIGKDD} Explorations}, 13(2):50--64, 2011.

\bibitem{TaoYL14}
W.~Tao, M.~Yu, and G.~Li.
\newblock Efficient top-k simrank-based similarity join.
\newblock {\em {PVLDB}}, 8(3):317--328, 2014.

\bibitem{Yu14}
W.~Yu, X.~Lin, and W.~Zhang.
\newblock Fast incremental simrank on link-evolving graphs.
\newblock In {\em ICDE}, pages 304--315, 2014.

\bibitem{Yu13}
W.~Yu, X.~Lin, W.~Zhang, L.~Chang, and J.~Pei.
\newblock More is simpler: Effectively and efficiently assessing node-pair
  similarities based on hyperlinks.
\newblock {\em PVLDB}, 7(1):13--24, 2013.

\bibitem{YuM15b}
W.~Yu and J.~A. McCann.
\newblock Efficient partial-pairs simrank search for large networks.
\newblock {\em {PVLDB}}, 8(5):569--580, 2015.

\bibitem{YuM15a}
W.~Yu and J.~A. McCann.
\newblock High quality graph-based similarity search.
\newblock In {\em {SIGIR}}, pages 83--92, 2015.

\bibitem{YZL12}
W.~Yu, W.~Zhang, X.~Lin, Q.~Zhang, and J.~Le.
\newblock A space and time efficient algorithm for simrank computation.
\newblock {\em World Wide Web}, 15(3):327--353, 2012.

\bibitem{ZhaoHS09}
P.~Zhao, J.~Han, and Y.~Sun.
\newblock P-rank: a comprehensive structural similarity measure over
  information networks.
\newblock In {\em {CIKM}}, pages 553--562, 2009.

\bibitem{ZhengZF0Z13}
W.~Zheng, L.~Zou, Y.~Feng, L.~Chen, and D.~Zhao.
\newblock Efficient simrank-based similarity join over large graphs.
\newblock {\em {PVLDB}}, 6(7):493--504, 2013.

\end{thebibliography}
\end{small}

\appendix
\section{Limitations of the Linearization Method}  \label{apnx:linear}

Recall that the linearization method \cite{MKK14} requires pre-computing the diagonal correction matrix $D$. Maehara et al.\ \cite{MKK14} prove that the diagonal elements in $D$ satisfy the following linear system:
\begin{equation} \label{eqn:exist-lsim-system}
\textrm{for all $k \in [1, n]$, } \qquad \sum_{\l=0}^{\infty}\sum_{i=1}^n c^\l {\left(p_{k,i}^{(\l)}\right)}^2 D(i,i) = 1,
\end{equation}
where $p_{k,i}^{(\l)}$ is the probability that $v_i$ is the $\l$-th step of a reverse random walk from $v_k$. Based on this, the linearization method estimates $p_{k,i}^{(\l)}$ with a set of reverse random walks, and then incorporates the estimated values into a truncated version of Equation~\eqref{eqn:exist-lsim-system}:
\begin{equation} \label{eqn:exist-lsim-system2}
\textrm{for all $k \in [1, n]$, } \qquad \sum_{\l=0}^{t}\sum_{i=1}^n c^\l {\left(\tilde{p}_{k,i}^{(\l)}\right)}^2 D(i,i) = 1,
\end{equation}
where $\tilde{p}_{k,i}^{(\l)}$ denotes the estimated version of $p_{k,i}^{(\l)}$. After that, it applies the Gauss-Seidel technique \cite{GV12} to solve Equation~\eqref{eqn:exist-lsim-system2}, and obtains an $n \times n$ diagonal matrix $\widetilde{D}$ that approximates $D$.

The above approach for deriving $\widetilde{D}$ is interesting, but it fails to provide any worst-case guarantee in terms of the pre-computation time and the accuracy of SimRank queries, due to the following reasons. First, because of the sampling error in $\widetilde{p}_{k,i}^{(\l)}$ and the truncation applied in Equation~\eqref{eqn:exist-lsim-system2}, $\widetilde{D}$ could differ considerably from $D$, which may in turn lead to significant errors in SimRank computation. There is no formal result on how large the error in $\widetilde{D}$ could be. Instead, Maehara et al.\ \cite{MKK14} only show that the error in $\widetilde{p}_{k,i}^{(\l)}$ can be bounded by using a sufficiently large sample set of reverse random walks; however, it does not translate into any accuracy guarantee on $\widetilde{D}$.

\begin{figure}[t]
\centering
\begin{small}
    \includegraphics[height=16mm]{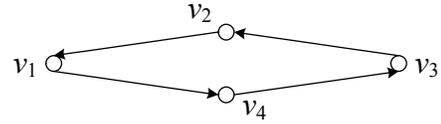}
    \figcapup
    \caption{An adversarial case for the linearization method.}
    \label{fig:apnx-linear-wrong}
\end{small}
\end{figure}

\begin{figure*}[htbp]
\begin{small}
\centering
\begin{tabular}{cccc}
\hspace{-4mm}\includegraphics[height=35.5mm]{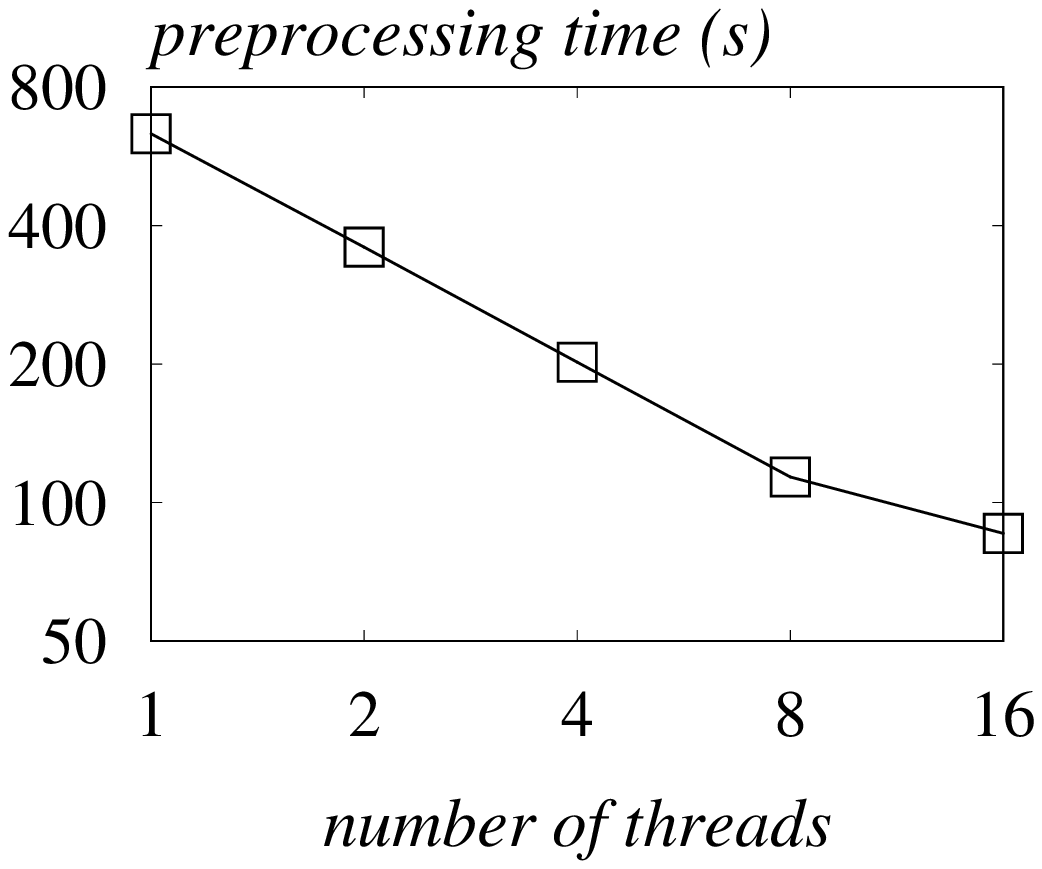}
&
\hspace{-10mm}\includegraphics[height=35.5mm]{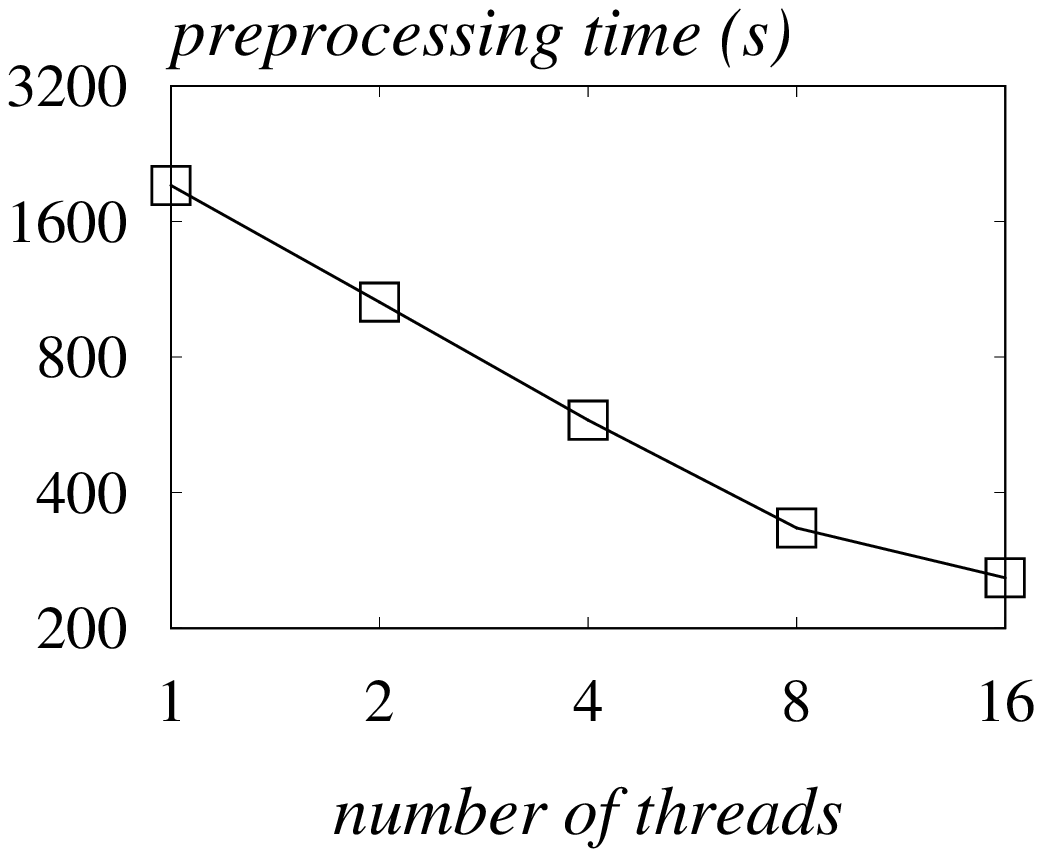}
&
\hspace{-10mm}\includegraphics[height=35.5mm]{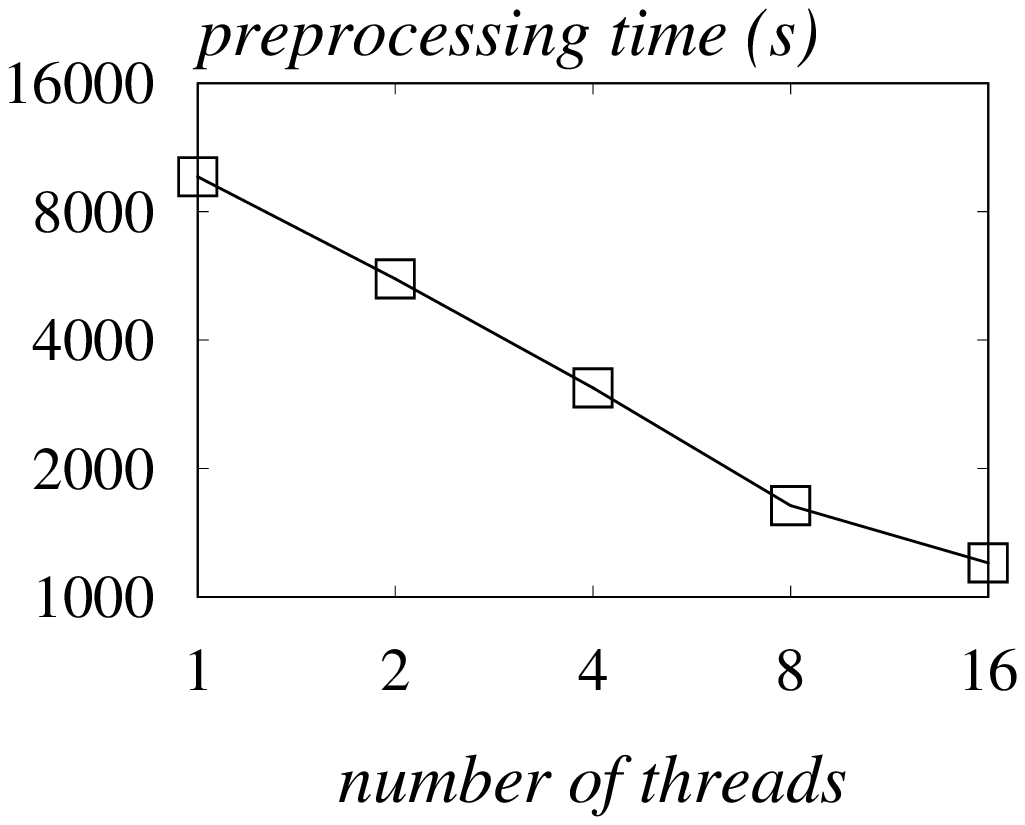}
&
\hspace{-10mm}\includegraphics[height=35.5mm]{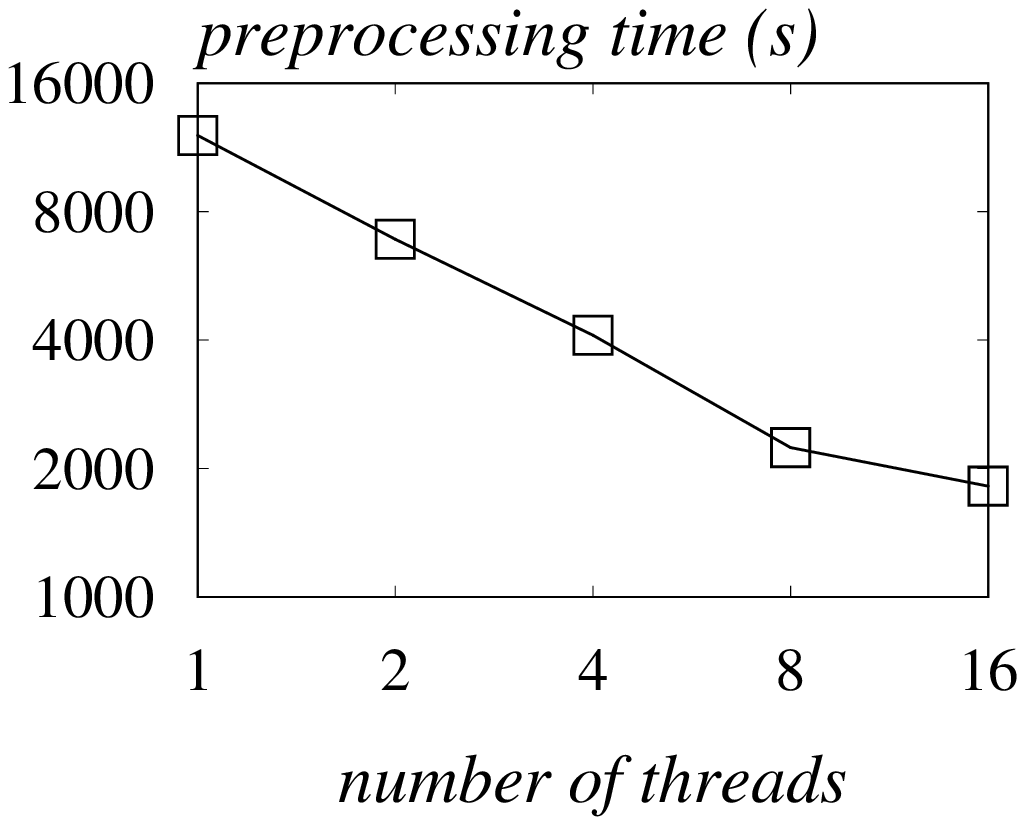}
\\
\hspace{-4mm} (a)  Google & \hspace{-4mm}(b) In-2004
& \hspace{-4mm}(c) LiveJournal & \hspace{-4mm}(d) Indochina
\end{tabular}
\figcapup  \vspace{-1mm} \caption{Preprocessing time vs.\ number of threads.} 
\label{fig:parallel}
\vspace{-0mm}
\end{small}
\end{figure*}

\begin{figure*}[htbp]
\begin{small}
\centering
\begin{tabular}{cccc}
\hspace{-5mm}\includegraphics[height=36mm]{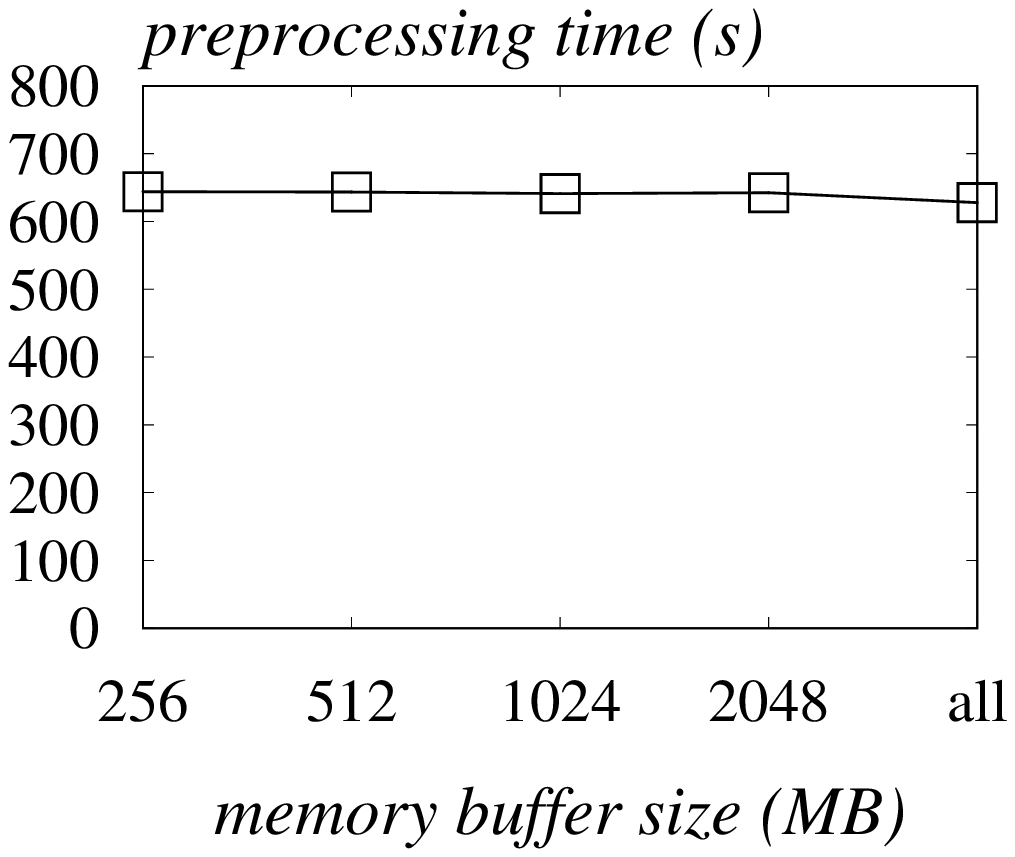}
&
\hspace{-11mm}\includegraphics[height=36mm]{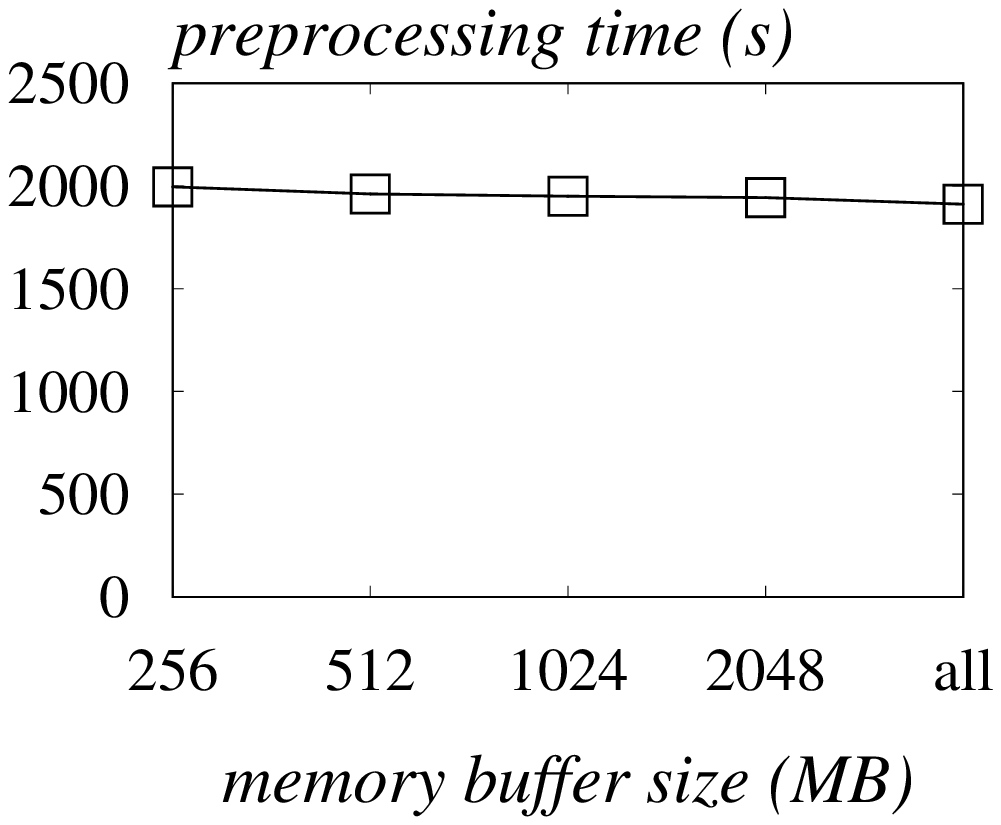}
&
\hspace{-11mm}\includegraphics[height=36mm]{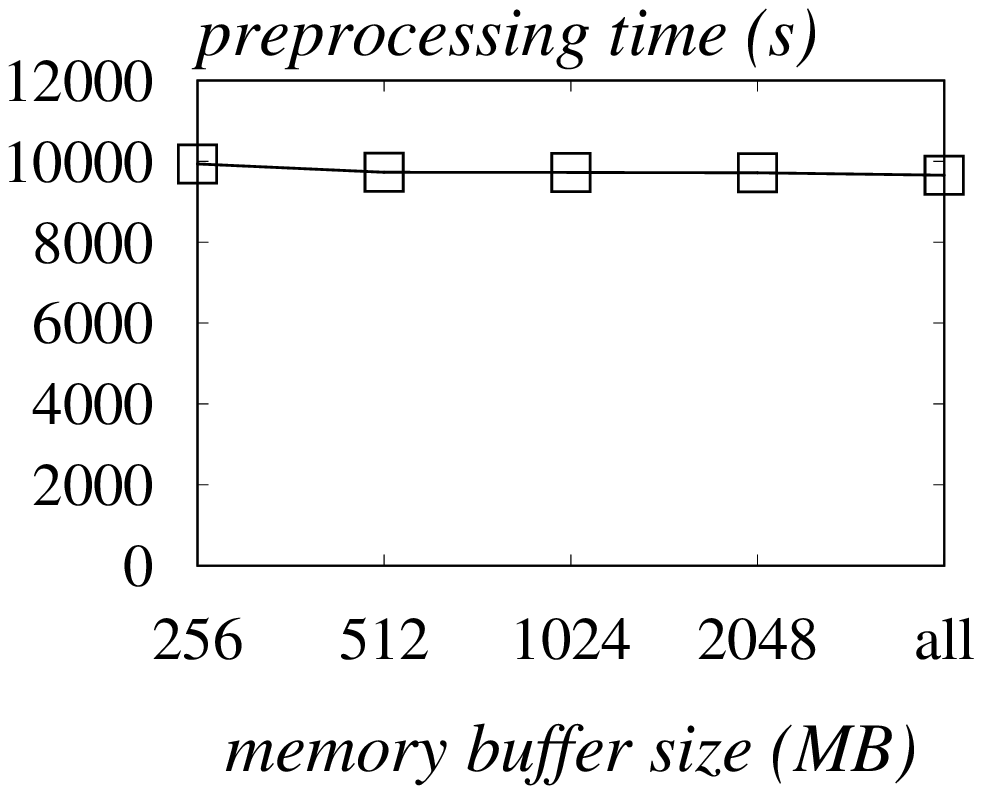}
&
\hspace{-11mm}\includegraphics[height=36mm]{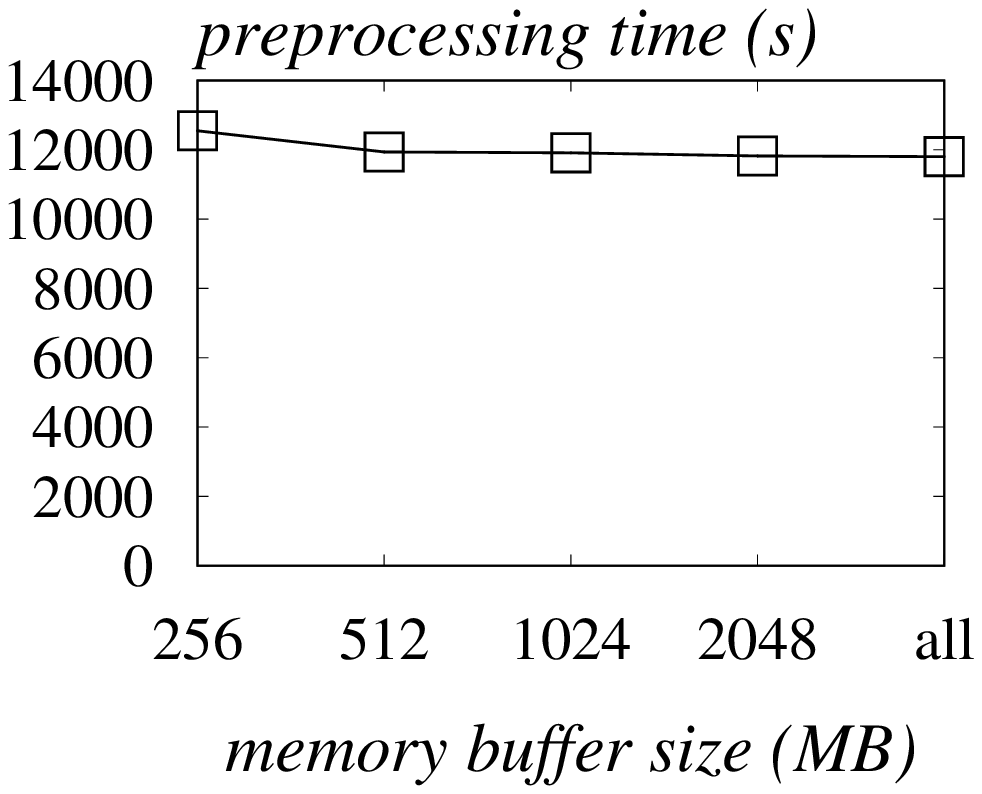}
\\
\hspace{-4mm} (a)  Google & \hspace{-4mm}(b) In-2004
& \hspace{-4mm}(c) LiveJournal & \hspace{-4mm}(d) Indochina
\end{tabular}
\figcapup \vspace{-2mm}  \caption{Preprocessing time vs.\ memory buffer size.} 
\label{fig:out-of-core}
\vspace{-0mm}
\end{small}
\end{figure*}

Second, even if $\widetilde{p}_{k,i}^{(\l)} = p_{k,i}^{(\l)}$ and Equation~\eqref{eqn:exist-lsim-system2} is not truncated, the Gauss-Seidel technique \cite{GV12} used by the linearization method to solve Equation~\eqref{eqn:exist-lsim-system2} may not converge. In particular, if we define an $n \times n$ matrix $M$ as
$$M(k, i) = \sum_{\l=0}^{\infty}\sum_{i=1}^n c^\l {\left(p_{k,i}^{(\l)}\right)}^2,$$
then the linearization method requires that $M$ should be diagonally dominant, i.e., for any $i$, $|M(i,i)| \ge \sum_{j\neq i} |M(i,j)|$. However, this requirement is not always satisfied. For example, consider the graph in Figure~\ref{fig:apnx-linear-wrong}. The linear system corresponding to the graph is

$$\dfrac{1}{1 - c^4}\left( \begin{array}{l l l l}
1   & c     & c^2   & c^3   \\
c^3 & 1     & c     & c^2   \\
c^2 & c^3   & 1     & c     \\
c   & c^2   & c^3   & 1     \\
\end{array}\right)
\left(\begin{array}{l}
D(1,1) \\ D(2,2) \\ D(3,3) \\ D(4,4)
\end{array}\right)
=
\left(\begin{array}{l}
1 \\ 1 \\ 1 \\ 1
\end{array} \right)$$
It can be verified the matrix $4\times4$ matrix $M$ on the left hand side is not diagonally dominant when $c=0.6$.

Finally, the number of iterations required by the Gauss-Seidel method is $O(\log \e^* / \log \rho)$, where $\e^*$ is the maximum error allowed in the solution to the linear system, and $\rho$ is the spectral radius of the iteration matrix used by the method \cite{GV12}. The value of $\rho$ depends on the input graph, and might be very close to $1$, in which case $\log \e^* / \log \rho$ can be an extremely large number.

\section{Hitting Probabilities vs.\ Personalized PageRanks} \label{apnx:pagerank}

Suppose that we start a random walk from a node $v_i$ following the outgoing edges of each node, with $1 - c_p$ probability to stop at each step. The probability that the walk stops at a node $v_j$ is referred to as the {\em personalized PageRank (PPR)} \cite{JehW03} from $v_i$ to $v_j$. PPR is well-adopted as a metric for measuring the {\em relevance} of nodes with respect to the input node $v_i$, and it has important applications in web search \cite{JehW03} and social network analysis \cite{GGL13}.

Our notion of hitting probabilities (HP) bears similarity to PPR, but differs in the following aspect:
\begin{enumerate}[topsep = 6pt, parsep = 6pt, itemsep = 0pt, leftmargin=20pt]
\item HP concerns the probability that the random walk reaches node $v_j$ at a particular step $\l$, but disregards whether the random walk stops at $v_j$;

\item PPR only concerns the endpoint $v_j$ of the random walk, and disregards all nodes before it.
\end{enumerate}

Our Algorithm~\ref{alg:backward} for computing approximate HPs is inspired by the {\em local update} algorithm \cite{ACL06,JehW03,FRCS05} proposed for computing approximate PPRs. Specifically, given a node $v_j$ and an error bound $\e$, the local update algorithm returns an approximate version of the PPRs from other nodes to $v_j$, with $\e$ worst-case errors. The algorithm starts by assigning a {\em residual} $1$ to $v_j$, and $0$ to any other node. Subsequently, the algorithm iteratively propagates the residual of each node to its in-neighbors, during which it computes the approximate PPR from each node to $v_j$. When the largest residual in all nodes is smaller than $\e$, the algorithm terminates. This algorithm is similar in spirit to our Algorithm~\ref{alg:backward}, but it cannot be directly applied in our context, due to the inherent differences between PPRs and HPs.

\section{Additional Experiments} \label{apnx:experiment}

In this section, we evaluate the parallel and out-of-core algorithms for constructing the index structures of {\em SLING} (presented in Section~\ref{sec:opt-parallel}), using the four largest datasets in Table~\ref{tbl:datasets}. First, we implement a multi-threaded version of {\em SLING}'s pre-computation algorithm, and measure its running time when the number of threads varies from $1$ to $16$ and all 64GB main memory on our machine is available. (The total number of CPU cores on our machine is $16$.) Figure~\ref{fig:parallel} illustrates the results. Observe that the algorithm achieves a near-linear speed-up as the number of threads increases, which is consistent with our analysis (in Section~\ref{sec:opt-parallel}) that {\em SLING} preprocessing algorithm is embarrassingly parallelizable.

Next, we implement an I/O-efficient version of {\em SLING}'s preprocessing algorithm, based on our discussions in Section~\ref{sec:opt-parallel}. Then, we measure the running time of the algorithm when it uses one CPU core along with a memory buffer of a pre-defined size. (We assume that the input graph is memory-resident, and we exclude it when calculating the memory buffer size.) Figure~\ref{fig:out-of-core} shows the processing time of the algorithm as the buffer size varies. Observe that the algorithm can efficiently process all tested graphs even when the buffer size is as small as $256$MB. In addition, the overhead of the algorithm does not increase significantly when the buffer size decreases, since the algorithm is CPU-bound. In particular, its only I/O cost is incurred by (i) writing each entry in the index once to the disk, and (ii) performing an external sort on the entries.

\section{Concentration Inequalities} \label{apnx:concen}

This section introduces the concentration inequalities used in our proofs. We start from the classic Chernoff bound.
\begin{lemma}[Chernoff Bound \cite{ChungL06}] \label{lmm:chernoff}
For any set $\{x_i\}$ ($i \in [1, n_x]$) of i.i.d.\ random variables with mean $\mu$ and $x_i \in [0, 1]$,
$$\Pr\left\{\left|\sum_{i=1}^{n_x} x_i - n_x \mu\right| \geq n_x \e\right\} \leq \exp\left(-\dfrac{n_x \cdot \e^2}{\frac{2}{3}\e + 2\mu}\right).$$
\end{lemma}

Our proofs also use a concentration bound on {\em martingales}, as detailed in the following.
\begin{definition}[Martingale]
A sequence of random variables $y_1, y_2, y_3, \cdots$ is a martingale if and only if $\mathbb{E}[y_i] < +\infty$ and $\mathbb{E}[y_{i+1}|y_1, y_2, \cdots, y_i] = y_i$ for any $i$. \done
\end{definition}

\begin{lemma}[\cite{ChungL06}] \label{lmm:martingale}
Let $y_1, y_2, y_3, \cdots$ be a martingale, such that $|y_1| \leq a$, $|y_{j+1} - y_j| \leq a$ for any $j \in [1, i - 1]$, and
$$\textrm{Var}[y_1] + \sum_{j=2}^i \textrm{Var}[y_j \mid y_1, y_2, \cdots, y_j - 1] \leq b_i,$$
where $\textrm{Var}[\cdot]$ denotes the variance of a random variable. Then, for any $\lambda > 0$,
$$\Pr\left\{y_i - \E[y_i] \geq \lambda\right\} \leq \exp\left(-\dfrac{\lambda^2}{\frac{2}{3}a\lambda + 2b_i}\right)$$
\end{lemma}

\section{Proofs} \label{apnx:proofs}

\vspace{-1mm}
\pheader
{\bf Proof of Lemma~\ref{lmm:our-scwalk}.} Let $s'(v_i, v_j)$ be the probability that $W_i$ and $W_j$ meet. If $v_i = v_j$, then $s'(v_i, v_j) = 1$, since $W_i$ and $W_j$ always meet at the first step. Suppose that $v_i \ne v_j$. Then, $s'(v_i, v_j)$ is the probability that $W_i$ and $W_j$ meet at or after the second step. Assume without loss of generality that the second steps of $W_i$ and $W_j$ are $v_k$ and $v_\l$, respectively. By definition, $s'(v_k, v_\l)$ equals the probability that $W_i$ and $W_j$ meet at or after $v_k$ and $v_\l$. Taking into account all possible second steps of $W_i$ and $W_j$, we have
\begin{align*}
s'(v_i, v_j) &= \sum_{v_k \in \inV(v_i), v_\l \in \inV(v_j)}\dfrac{\sqrt{c}}{|\inV(v_i)|} \cdot
                \dfrac{\sqrt{c}}{|\inV(v_j)|} \cdot
                s'(v_k, v_\l) \\
         &= \dfrac{c}{|\inV(v_i)|\cdot |\inV(v_j)|} \sum_{v_k \in \inV(v_i), v_\l \in \inV(v_j)} s'(v_k, v_\l).
\end{align*}
As such, $s'(v_i, v_j)$ have the same definition as $s(v_i, v_j)$ (see Equation~\eqref{eqn:def-sim}), which indicates that $s'(v_i, v_j) = s(v_i, v_j)$. \done

\pheader
{\bf Proof of Lemma~\ref{lmm:our-lsim}.} First, we define the following events:
\begin{itemize}
\item $E(v_i, v_j)$ : Two $\sqrt{c}$-walks starting from $v_i$ and $v_j$, respectively, meet each other.
\item $L(v_i, v_j, v_k, \l)$ : Two $\sqrt{c}$-walks starting from $v_i$ and $v_j$, respectively, {\em last meet} each other at the $\l$-th step at $v_k$.
\end{itemize}
As we discuss in Section~\ref{sec:our-overview}, two different events $L(v_i, v_j, v_k, \l)$ and $L(v_i, v_j, v_k', \l')$ are mutually exclusive whenever $v_k \ne v_k'$ or $\l \ne \l'$. Therefore,
$$\Pr\{E(v_i, v_j)\} = \sum_{\l=0}^{+\infty} \sum_{k =1}^{n} \Pr\{L(v_i, v_j, v_k, \l)\}$$

Observe that the probability of $L(v_i, v_j, v_k, \l)$ can be computed by multiplying the following two probabilities:
\begin{enumerate}[topsep = 6pt, parsep = 6pt, itemsep = 0pt, leftmargin=20pt]
\item The probability that two $\scw$-walks $W_i$ and $W_j$ from $v_i$ and $v_j$, respectively, meet at $v_k$ at step $\l$.

\item Given that $W_i$ and $W_j$ meet at $v_k$ step $\l$, the probability that they do not meet at steps $\l+1, \l+2, \ldots$.
\end{enumerate}
The first probability equals $h^{(\l)}(v_i, v_k) \cdot h^{(\l)}(v_i, v_k)$. Meanwhile, since the $(x+1)$-th step of any $\scw$-walk depends only on its $x$-th step, the second probability should equal the probability that two $\scw$-walks from $v_k$ never meet after the $0$-th step, which in turn equals $d_k$. Hence, we have
\begin{align*}
s(v_i, v_j) & = \Pr\{E(v_i, v_j)\} = \sum_{\l=0}^{+\infty} \sum_{k =1}^{n} \Pr\{L(v_i, v_j, v_k, \l)\} \\
&= \sum_{\l=0}^{+\infty} \sum_{k=1}^{n} \left(h^{(\l)}(v_i, v_k) \cdot d_k \cdot h^{(\l)}(v_j, v_k)\right),
\end{align*}
which completes the proof. \done

\pheader
{\bf Proof of Lemma~\ref{lmm:our-dk-D}.} Let ${R = \sqrt{c} \cdot P}$, and $R^{\l}$ be the $\l$-th power of $R$. We have
$$ R^{0}(k,i) =
\begin{cases}
1 & \textrm{if $i = k$} \\
0 & \textrm{otherwise}
\end{cases}$$
Hence, $R^{0}(k, i) = h^{(0)}(v_i, v_k)$ for all $v_i$ and $v_k$. Assume that for a certain $\l$, we have ${R^{\l}(k, i) = h^{(\l)}(v_i, v_k)}$ for all $v_i$ and $v_k$. Then,
\begin{align*}
R^{\l + 1}(k,i) &= \left(\sqrt{c}\cdot P \cdot R^{\l}\right)(k,i) \\
&= \sum_{j=1}^{n} \left(\sqrt{c} \cdot P(k,j) \cdot R^{\l}(j,i)\right) \\
&= \sum_{\textrm{each out-neighbor $v_j$ of $v_k$}}\left(\frac{\sqrt{c}}{|\inV(v_j)|} \cdot h^{(\l)}(v_i, v_j)\right) \\
&= h^{(\l+1)}(v_i, v_k).
\end{align*}
Therefore, $R^{\l}(k, i) = h^{(\l)}(v_i, v_k)$ for all $v_i$, $v_k$, and $\l$.

Let $\tilde{D}$ be the $n \times n$ diagonal matrix whose $k$-th diagonal element is $d_k$. Then, Equation \eqref{eqn:our-lsim} can be written as:
$$S = \sum_{\l=0}^{+\infty} \left(\left(R^{\l}\right)^{\top}\tilde{D}R^{\l}\right)$$
By multiplying $R^{\top}$ and $R$ on the left and right, respectively, on both side of the equation, we have
$$S = R^{\top} S R + \tilde{D} = cP^{\top}SP + \tilde{D}.$$
This indicates that $\tilde{D}$ is a diagonal correction matrix. Since the diagonal correction matrix is unique \cite{MKK14}, we have $\tilde{D} = D$. \done

\balance

\pheader
{\bf Proof of Lemma~\ref{lmm:our-D}.}
By the Chernoff Bound in Lemma~\ref{lmm:chernoff},
\begin{align*}
\Pr\left\{\left|\frac{cnt}{n_r} - \mu \right| \geq \e/c\right\} 
&\leq 2\exp\left(-\dfrac{n_r (\e/c)^2}{\frac{2}{3}\e/c + 2\mu}\right) \\
&= 2\exp\left(-\dfrac{\frac{2}{3}\e/c + 2}{\frac{2}{3}\e/c + 2\mu} \log{\dfrac{2}{\delta_d}}\right) \\
&= \delta_d
\end{align*}
Therefore, $|\tilde{d_k} - d_k| = c \cdot |\frac{cnt}{n_r} - \mu| \leq \e$ with at least $1 - \delta_d$ probability. \done

\pheader
{\bf Proof of Lemma~\ref{lmm:our-R}.}
According to Algorithm \ref{alg:backward}, for all $\tilde{h}^{(\l)}(v_i, v_j) \in H(v_i)$, we have
$$\theta \le \tilde{h}^{(\l)}(v_i, v_j) \leq h^{(\l)}(v_i, v_j).$$
Then, for each node $v_i$ and each step $\l$,
$$\sum_{v_j \in V} \tilde{h}^{(\l)}(v_i, v_j) \leq \sum_{v_j \in V} h^{(\l)}(v_i, v_j) = \sqrt{c}^{\l}$$
Therefore, there are at most $(\sqrt{c})^{\l}/\theta$ nodes $v_j$ such that $\tilde{h}^{(\l)}(v_i, v_j) \in H(v_i)$. Therefore, the size of $H(v_i)$ is
$$|H(v_i)| \; \leq \; \sum_{\l=0}^{+\infty} \dfrac{\left(\sqrt{c}\right)^{\l}}{\theta} \; = \; O(1/\theta)$$
Let $\bar{d}$ be the average out-degree of the $G$. Since a local update is performed on each entry $\tilde{h}^{(\l)}(v_i, v_j) \in H(v_i)$, the running time of the algorithm is $\O(\bar{d}n/\theta) = \O(m/\theta)$.

Let $\e_{\l}$ be the upper bound of $|\tilde{h}^{(\l)}(v_i, v_j) - h^{(\l)}(v_i, v_j)|$ for all nodes $v_i$ and $v_j$ at step $\l$. When $\l = 0$, we have $\e_{0} \leq \dfrac{1 - \left(\sqrt{c}\right)^{0}}{1 - \sqrt{c}}\theta$. Assume that $\e_{\l} \leq \dfrac{1 - \left(\sqrt{c}\right)^{\l}}{1 - \sqrt{c}}\theta$ holds for a certain $\l$. Then,
\begin{align*}
\e_{\l+1} &\leq  \sqrt{c} \cdot \e_{\l} + \theta = \dfrac{1 - \sqrt{c}^{\l+1}}{1 - \sqrt{c}} \theta.
\end{align*}
Thus, the lemma is proved. \done

\pheader
{\bf Proof of Lemma~\ref{lmm:our-error-allowed}.}
Given that $\left| \tilde{d}_k - d_k \right| \leq \e_d$ and ${-\e_h^{(\l)} \leq \tilde{h}^{(\l)}(v_k, v_x) - h^{(\l)}(v_k, v_x) \leq 0}$ for any $k, x, \l$, we have
\begin{align*}
& \tilde{h}^{(\l)}(v_i, v_k) \cdot \tilde{d} \cdot \tilde{h}^{(\l)}(v_j, v_k) - h^{(\l)}(v_i, v_k) \cdot d_k \cdot h^{(\l)}(v_j, v_k) \\
& \leq h^{(\l)}(v_i, v_k)\cdot \tilde{d_k} \cdot h^{(\l)}(v_j, v_k)  - h^{(\l)}(v_i, v_k)\cdot d_k \cdot h^{(\l)}(v_j, v_k) \\
& \leq h^{(\l)}(v_i, v_k)\cdot \e_d \cdot h^{(\l)}(v_j, v_k).
\end{align*}
Therefore,
\begin{align*}
\tilde{s}(v_i, v_j) - s(v_i, v_j) & \le \sum_{\l=0}^{+\infty} \sum_{k=1}^{n}  h^{(\l)}(v_i, v_k)\cdot \e_d \cdot h^{(\l)}(v_j, v_k)\\
&\leq \sum_{\l=0}^{+\infty} \e_d \cdot (\sqrt{c})^{\l} \cdot (\sqrt{c}^{\l}) \quad = \dfrac{\e_d}{1-c}.
\end{align*}
Meanwhile,
\begin{align*}
& h^{(\l)}(v_i, v_k) \cdot d_k \cdot h^{(\l)}(v_j, v_k) - \tilde{h}^{(\l)}(v_i, v_k) \cdot \tilde{d} \cdot \tilde{h}^{(\l)}(v_j, v_k) \\
&\leq h^{(\l)}(v_i, v_k)\cdot d_k \cdot h^{(\l)}(v_j, v_k) \\
& \qquad - \left(h^{(\l)}(v_i, v_k) - \e_h^{(\l)}\right)\cdot\left(d_k - \e_d\right)\cdot\left(h^{(\l)}(v_j, v_k) - \e_h^{(\l)}\right) \\
&=  \left(h^{(\l)}(v_i, v_k) - \e_h^{(\l)}\right) \cdot \e_d \cdot \left(h^{(\l)}(v_j, v_k) - \e_h^{(\l)}\right) \\
& \qquad + \e_h^{(\l)} \cdot d_k \cdot \left(h^{(\l)}(v_j, v_k) - \e_h^{(\l)}\right) + \e_d \cdot \e_h^{(\l)} \cdot h^{(\l)}(v_i, v_k) \\
\end{align*}
Hence,
\begin{align*}
& s(v_i, v_j) - \tilde{s}(v_i, v_j) \\
&= \sum_{\l=0}^{+\infty} \sum_{k=1}^{n} \bigg(\left(h^{(\l)}(v_i, v_k) - \e_h^{(\l)}\right) \cdot \e_d \cdot \left(h^{(\l)}(v_j, v_k) - \e_h^{(\l)}\right) \\
& \qquad + \e_h^{(\l)} \cdot d_k \cdot \left(h^{(\l)}(v_j, v_k) - \e_h^{(\l)}\right) + \e_d \cdot \e_h^{(\l)} \cdot h^{(\l)}(v_i, v_k)\bigg) \\
&\leq \sum_{\l=0}^{+\infty} \left((\sqrt{c})^{\l} \cdot \e_d \cdot  (\sqrt{c}^{\l}) + 2 \cdot \e_h^{(\l)} \cdot (\sqrt{c})^{\l}\right)\\
&= \dfrac{\e_d}{1-c} + 2 \sum_{\l=0}^{+\infty} \left((\sqrt{c}^{\l})\cdot \e_h^{(\l)} \right).
\end{align*}
This completes the proof. \done

\pheader
{\bf Proof of Theorem~\ref{thm:err-bound}.}
By Lemma~\ref{lmm:our-R}, for all $k, x, l$,
$$-\dfrac{1 - (\sqrt{c})^{\l}}{1 - \sqrt{c}}\theta \leq \tilde{h}^{(\l)}(v_k, v_x) - h^{(\l)}(v_k, v_x) \leq 0.$$
Then,
\begin{align*}
& \dfrac{\e_d}{1-c} + 2\sum_{\l=0}^{+\infty} \left( (\sqrt{c})^{\l}\cdot \e_h^{(\l)} \right) \\
& = \dfrac{\e_d}{1-c} + 2\sum_{\l=0}^{+\infty} \left( (\sqrt{c})^{\l} \cdot \dfrac{1 - (\sqrt{c})^{\l}}{1 - \sqrt{c}}\theta \right) \\
&= \dfrac{\e_d}{1-c} + \dfrac{2\sqrt{c}}{(1-c)(1-\sqrt{c})}\theta.
\end{align*}
By Lemma \ref{lmm:our-D}, $\left|\tilde{d}_k - d_k\right| \leq \e_d$ holds with at least $1 - \delta_d$ probability. Since $\delta_d \leq \delta / n$, with at least $1 - \delta$ probability,
$$\left|\tilde{d}_k - d_k\right| < \e_d \textrm{ , for all $k$}$$
Therefore, By Lemma \ref{lmm:our-error-allowed}, $|\tilde{s}(v_i, v_j) - s(v_i, v_j)| \leq \e$ holds with at least $1 - \delta$ probability. \done

\pheader
{\bf Proof of Lemma~\ref{lmm:opt-D-error}.} Let $\e^* = \e_d / c$. Then $n_r = \dfrac{14}{3\e^*}\log{\dfrac{4}{\delta_d}}$, $n_r^* = \dfrac{2\mu^* + \frac{2}{3}\e^*}{{\e^*}^2}\log{\dfrac{4}{\delta_d}}$, and $\mu^* = \hat{\mu} + \sqrt{\hat{\mu}\e^*}$.
In the second part of the sampling procedure, if $\mu^* \geq \mu$, then by Lemma~\ref{lmm:martingale},
\begin{align*}
\Pr\{|\tilde{\mu} - \mu| > \e^*\}
    &\leq 2\exp\left(-\dfrac{n_r^* {\e^*}^2}{2\mu + \frac{2}{3}\e^*}\right) \\
    &= 2\exp\left(-\dfrac{2\mu^* + \frac{2}{3}\e^*}{2\mu + \frac{2}{3}\e^*}\log{\dfrac{4}{\delta_d}}\right) \quad \leq \dfrac{\delta_d}{2}.
\end{align*}
We differentiate two cases: $\mu \ge 2\e^*$ and $\mu < 2\e^*$. Assume that $\mu \ge 2\e^*$. Then,
\begin{align*}
\Pr\{\hat{\mu} \leq \mu/2\} &\leq  \exp\left(-\dfrac{n_r (\mu/2)^2}{2\mu+\frac{2}{3}\cdot \mu/2}\right) = \exp\left(-\dfrac{n_r \mu}{28/3}\right) \\
    &\leq  \exp\left(-\dfrac{\frac{14}{3\e}\log{\frac{4}{\delta_d}} \cdot 2\e}{28/3}\right) =  \dfrac{\delta_d}{4}
\end{align*}
Then, given $\mu > 2\e$ and $\hat{\mu} \geq \mu/2 > \e^*$,
\begin{align*}
& \Pr\{\mu > \hat{\mu} + \sqrt{\hat{\mu}\e^*}\} \\
& \leq \exp\left(-\dfrac{n_r \hat{\mu} \e^*}{2 \mu + \frac{2}{3}\sqrt{\hat{\mu}\e^*}}\right) \quad \leq \exp\left(-\dfrac{n_r \hat{\mu} \e}{4 \hat{\mu} + \frac{2}{3} \hat{\mu}}\right) \\
& = \exp\left(-\dfrac{n_r \e^*}{14/3}\right) \quad = \exp\left(-\dfrac{\frac{14}{3\e^*}\log{\frac{4}{\delta_d}} \cdot \e^*}{14/3}\right) \quad \leq \dfrac{\delta_d}{4}.
\end{align*}
Therefore, $\mu$ is estimated with at most $\e^*$ error with at least ${1 - (\frac{\delta}{4} + \frac{\delta}{4} + \frac{\delta}{2}) = 1 - \delta}$ probability.

Now consider that $\mu < 2\e^*$. We have
\begin{align*}
& \Pr\{|\mu - \hat{\mu}| \geq \e^*\} \leq 2\exp\left(-\dfrac{n_r {\e^*}^2}{2\mu + \frac{2}{3}\e^*}\right) \\
&\leq 2\exp\left(-\dfrac{n_r {\e^*}^2}{4\e^* + \frac{2}{3}\e^*}\right)  \; = 2\exp\left(-\dfrac{\frac{14}{3\e^*}\log{\frac{4}{\delta_d}} \cdot \e^*}{14/3}\right) \; = \dfrac{\delta_d}{2}.
\end{align*}
If $\hat{\mu} < \e^*$, then $\mu$ is estimated with at most $\e^*$ error with at least $1 - \delta_d/2$ probability. On the other hand, if $\hat{\mu} > \e^*$ ,then with at least $1 - \delta_d/2$ probability,
$$\mu < \hat{\mu} + \e^* < \hat{\mu} + \sqrt{\hat{\mu}\e^*}.$$
Therefore, $\mu$ is estimated with at most $\e^*$ error with at least ${1 - (\frac{\delta_d}{2} + \frac{\delta_d}{2}) = 1 - \delta_d}$ probability.

In summary, with at least $1 - \delta_d$ probability, $\mu$ is estimated with at most $\e^* = \e_d / c$ error, in which case $d_k$ is estimated with at most $\e_d$ error. \done

\pheader
{\bf Proof of Lemma~\ref{lmm:opt-D-time}.} Let $\e^* = \e_d / c$. Then, $n_r = \dfrac{14}{3\e^*}\log{\dfrac{4}{\delta_d}}$, $n_r^* = \dfrac{2\mu^* + \frac{2}{3}\e^*}{{\e^*}^2}\log{\dfrac{4}{\delta_d}}$, and $\mu^* = \hat{\mu} + \sqrt{\hat{\mu}\e^*}$. By Lemma~\ref{lmm:martingale}, for any $k \geq 1$,
\begin{align*}
&\Pr\Big\{k \cdot (\mu + \e^*) \leq \hat{\mu} - \mu \leq (k+1) \cdot (\mu + \e^*)\Big\} \\
& \le \Pr\big\{\hat{\mu} - \mu \geq k(\mu + \e^*)\big\} \leq \exp\left(-\dfrac{k^2 \cdot (\mu + \e^*)^2 \cdot n_r}{2\mu + \frac{2}{3} (\mu + \e^*)}\right) \\
& \le \exp\left(-\dfrac{k^2\cdot (\mu+\e^*)^2 \cdot n_r}{\frac{8}{3}\cdot k \cdot (\mu+\e^*)}\right) \leq \exp\left(-k\log{\dfrac{4}{\delta_d}}\right) \leq \left(\dfrac{1}{4}\right)^k
\end{align*}
Then, we have the following upper bound on the expectation of $\hat{\mu}$:
\begin{align*}
\mathbb{E}[\hat{\mu}] &\leq (2\mu + \e^*) \cdot \Pr\{\hat{\mu} \leq 2\mu + \e^*\} + \sum_{k=1}^{+\infty} \dfrac{(k+1)(\mu + \e^*)}{4^k} \\
&\leq 2\mu + \e^* + \sum_{k=1}^{+\infty} \dfrac{(k+1)(\mu + \e^*)}{4^k} \quad = \dfrac{25\mu + 16\e^*}{9}
\end{align*}
Hence, we have the following upper bound on the the expected number of $\sqrt{c}$-walk pairs needed:
\begin{align*}
&\dfrac{14}{3\e^*}\log{\dfrac{4}{\delta_d}} + \dfrac{2 \cdot \mathbb{E}[\mu^*] + \frac{2}{3}\e^*}{{\e^*}^2} \log{\dfrac{4}{\delta_d}} \\
& \le \dfrac{14}{3\e^*}\log{\dfrac{4}{\delta_d}} + \dfrac{2 \cdot\mathbb{E}[\hat{\mu} + (\hat{\mu} + \e^*)] + \frac{2}{3}\e^*}{{\e^*}^2} \log{\dfrac{4}{\delta_d}} \\
& = \O\left(\frac{\mu + \e_d}{\e_d^2} \log{\frac{1}{\delta_d}}\right)
\end{align*}
Let $\l_i$ be the length of the shorter one of the $i$-th pair of $\sqrt{c}$-walks. Then $\l_i$ is identically geometrically distributed with success probability $1 - \sqrt{c}$. Then the upper bound on the expected running time of the algorithm is given by
\begin{align*}
\mathbb{E}[\sum_{i=1}^{n_r^*} \l_i] &= \mathbb{E}[n_r^*] \cdot \mathbb{E}[\l_i] = \O\left(\frac{\mu + \e_d}{\e_d^2} \log{\frac{1}{\delta_d}}\right)
\end{align*} \done

\pheader
{\bf Proof of Lemma~\ref{lmm:opt-D-optimal}.} The $\mathcal{A}^*$ be the algorithm $\mathcal{A}$ defined in the end of Section 5.1, except that it returns an estimation $\tilde{\mu_z}$ with a relative error guarantee, i.e., with at lesat $1-\delta$ probability,
$$(1-\lambda) \cdot \mu_Z \; \leq \; \tilde{\mu_Z} \; \leq \; (1+\lambda) \cdot \mu.$$
The {\em lower bound theorem} in \cite{DKLR00} shows that the expected number of samples taken by $\mathcal{A}^*$ is
$$\Omega\left(\frac{\max\{\mu_z (1-\mu_z), \lambda \mu_z\}}{\lambda^2 \cdot \mu_z^2}\log\frac{1}{\delta}\right).$$

Observe that if $\lambda = \e/\mu_z$, then $\mathcal{A}^*$ ensures at most $\e$ additive error. In that case, if $\mu<0.5$, the expected number of samples taken by $\mathcal{A}^*$ is
\begin{align*}
&\Omega\left(\frac{\max\{\mu_z (1-\mu_z), \lambda \mu_z\}}{\lambda^2 \cdot \mu_z^2}\log\frac{1}{\delta}\right) = \Omega\left(\frac{\max\{\mu_z, \e\}}{\e^2}\log\frac{1}{\delta}\right),
\end{align*}
which completes the proof. \done

\pheader
{\bf Proof of Lemma~\ref{lmm:ext-single-source}.}
For given $\l$ and $t$, consider two random walks $W_1$ and $W_2$, such that $W_1$ starts from $v_i$ at time $0$, while $W_2$ starts at $v_j$ at time $\l - t$. Consider the probability that $W_1$ and $W_2$ meet at time $\l$ and do not meet again, denoted by $\hat{\rho}_{\l}^{(t)}(v_j)$. Then, we have
\begin{itemize}
\item $\hat{\rho}_{\l}^{(0)}(v_k) = h^{(\l)}(v_i, v_k) \cdot d_k$, and
\item $\hat{\rho}_{\l}^{(t+1)}(v_j) = \dfrac{\sqrt{c}}{|\inV(v_j)|} \sum_{v_k\in \inV(v_j)} \hat{\rho}_{\l}^{(t)}(v_k)$.
\end{itemize}
Observe that, in Algorithm \ref{alg:single-source}, if we ignore the thresholding approximation and the errors in $\tilde{h}^{(\l)}(v_i, v_k)$, then the algorithm exactly corresponds to the iterative process defined by the above equations. Moreover, $\hat{\rho}_{\l}^{(\l)}(v_j)$ is the probability that two random walks $W_1$ and $W_2$ starting at $v_i$ and $v_j$ together meet each other after $\l$ steps, for the last time, which exactly equals $\sum_{k=1}^{n} h^{(\l)}(v_i, v_k) \cdot d_k \cdot h^{(\l)}(v_j, v_k)$. Therefore,
$$s(v_i, v_j) = \sum_{\l = 0}^{+\infty} \hat{\rho}_{\l}^{(\l)}(v_j).$$

Now, we consider the error in each $\hat{\rho}_{\l}^{(t)}(v_j)$. Let $\e_{\l}^{(t)}$ be the error at the $t$-th step. Then,
\begin{itemize}
\item For $t=0$, $\e_{\l}^{(0)} = |\hat{\rho}_{\l}^{(0)}(v_k) - \rho_{\l}^{(0)}| \leq \sqrt{c}^{\l} \e_d + \dfrac{1 - \sqrt{c}^{\l}}{1 - \sqrt{c}} \cdot \theta$.
\item $\e_{\l}^{(t+1)} \leq \sqrt{c} \cdot \e_{\l}^{(t)} + \sqrt{c}^{\l}\theta$
\end{itemize}
By solving the inequality, we get
$$\e_{\l} \leq c^{\l}\e_d + \dfrac{2 - 2\sqrt{c}^{\l}}{1 - \sqrt{c}}\theta$$
Therefore, the total error of the algorithm is
\begin{align*}
& \sum_{\l=0}^{+\infty} \left(c^{\l}\e_d + \dfrac{2 - 2\sqrt{c}^{\l}}{1 - \sqrt{c}}\theta\right) = \dfrac{\e_d}{1-c} + \dfrac{2\sqrt{c}}{(1-c)(1-\sqrt{c})}\theta \; \leq \e
\end{align*}

For each $\l$ and $t$, the calculation of $\rho_{\l}^{(t+1)}$ requires at most $O(m)$ times (i.e., scanning all the edges in the worst case). Since all entries in $H(v_i)$ are greater than $\theta$, $\l$ is at most $\log{\frac{1}{\theta}}$. Therefore, the total running time of the algorithm if bounded by
$$O\left(\sum_{\l = 0}^{\log(1/\theta)} \l \cdot m\right) = \O\left(m\log^2\frac{1}{\e}\right).$$

\end{sloppy}
\end{document}